\documentstyle[11pt,aaspp4]{article}

\def\cf{{\it cf. }}
\def\etal{{\it et al. }}
\def\eg{{\it e.g. }}

\def\beginmath{\begin{displaymath}}
\def\endmath{\end{displaymath}}

\def\begineq{\begin{equation}}
\def\endeq{\end{equation}}

\def\beginmlet{\begin{mathletters}}
\def\endmlet{\end{mathletters}}

\slugcomment{Submitted to ApJ}

\lefthead{Wan, Daly, \& Guerra}
\righthead{Key Parameters of FRII Radio Sources}

\begin{document}

\title{Empirical Determinations of Key Physical Parameters 
Related to Classical Double Radio Sources}

\author{Lin Wan, Ruth A. Daly\altaffilmark{1,2}, and Erick J. Guerra\altaffilmark{3}}
\affil{Department of Physics, Joseph Henry Laboratories, Princeton University, 
    Princeton, NJ 08544}

\altaffiltext{1}{Present address: Department of Physics, Bucknell University,
Lewisburg, PA 17837; daly@bucknell.edu}
\altaffiltext{2}{National Young Investigator}
\altaffiltext{3}{Present address: Department of Chemistry and Physics,
Rowan University, Glassboro, NJ 08028;\\ eddie@scherzo.rowan.edu}

\begin{abstract}

Multi-frequency radio observations of the radio bridge of powerful
classical double radio sources can be used to determine: the beam
power of the jets emanating from the AGN; the total time the
source will actively produce jets that power large-scale
radio emission; the thermal pressure of the medium in the
vicinity of the radio source; and the total mass, including dark
matter, of the galaxy or cluster of galaxies traced by the
ambient gas that surrounds the radio source.  The theoretical
constructs that allow a determination of each of these quantities
using radio observations are presented and discussed.  
Empirical determinations of each of these quantities are obtained
and analyzed.  

A sample of 14 radio galaxies and 8 radio loud quasars with
redshifts between zero and two for which there is enough radio
information to be able to determine the physical
parameters listed above was studied in detail.  
The beam power was determined for each beam of each AGN (so there
are two numbers for each source), and these AGN are
highly symmetric in terms of beam powers.  Typically, the  
beam powers are about $10^{45}\, \mbox{erg s}^{-1}$.
No strong correlation is seen
between the beam power and the core - hot spot separation, 
which suggests that the beam power is roughly constant
over the lifetime of a source. The beam power increases
with redshift, which is significant after excluding 
correlations between the radio power and redshift. 
The relationship between the beam power and
the radio power is not well constrained by the current data.

The characteristic or total time the AGN will actively produce a
collimated outflow is determined.  Typical total lifetimes are
$\sim 10^7$ years.  Typical source lifetimes decrease with
redshift, and this relation can explain the decrease
in the average source size (hot spot - hot spot separation) with
redshift.  Thus, high-redshift sources are smaller because they
have shorter lifetimes, since the rate of growth of sources is found to
increase with redshift.

A new method of estimating the thermal pressure of the ambient
gas in the vicinity of a powerful classical double radio 
source is presented. This new estimate is independent of 
synchrotron and inverse Compton aging arguments, and depends
only upon the properties of the radio lobe and the shape of the
radio bridge, which is used to determine the Mach number of lobe
advance.  A detailed radio map of the radio bridge at a single
radio frequency can be used to estimate the thermal pressure of
the ambient gas. 
Thermal pressures on the order of $10^{-10}\, \mbox{dyne cm}^{-2}$, 
typical of gas in low-redshift clusters of galaxies, are found for
the environments of the sources studied here. 
It is shown that
appreciable amounts of cosmic microwave background
diminution are expected from
many of these clusters, which could be observable at high frequency
where the emission from the radio sources is weak.

The total gravitational mass of the host cluster of galaxies is 
estimated using the composite pressure profile and 
the equation of hydrostatic equilibrium for cluster
gas. Masses and total density profiles very similar to those
of low-redshift clusters of galaxies are obtained.  Thus, 
some clusters of galaxies exist at redshifts
of two or so.  
The redshift evolution of the cluster mass is not well
determined at present.  The current data do not indicate any negative
evolution of the cluster mass, contrary to what is expected in a
high density universe. 

\end{abstract}


\section{\label{sec:intro}Introduction}

Classical double radio sources, also known
as FRII radio sources (Fanaroff \& Riley, 1974), have been
extensively studied.  Numerous surveys of
radio sources as well as detailed studies of individual sources 
have provided a wealth of radio data on many sources. Theoretical models 
for the sources developed and discussed by various authors (\eg Blandford
\& Rees, 1974; Scheuer, 1974, 1982; Begelman, Blandford, \& Rees,
1984; Begelman \& Cioffi, 1989; Daly 1990) lend much understanding to the
physics of the sources. It is believed that FRII sources are
powered by highly collimated outflows from an active galactic
nucleus (AGN), and that the radio emission from these sources is
the result of an interaction between the beam, or jet, and the
ambient medium.  As a result, careful study of FRII sources 
can yield very useful information 
about the FRII sources, their environments, and
their central engines. 

Daly (1994, 1995) presents a model for powerful 
extended FRII sources and showed
how key parameters of an FRII source and its environment, such as
the beam power and the ambient gas density, 
can be estimated using multi-frequency radio data. 
Results for a sample of powerful 3CR radio sources are also
presented by the same author (Daly 1994, 1995). 
A larger sample was later compiled by Daly and collaborators and
is studied in detail.  Many results from these studies have 
been presented in a series of papers (Wellman \& Daly 1996a,b;
Wellman 1997; Wellman, Daly, \& Wan 1997a, b [hereafter WDW97a,b]; 
Wan \& Daly 1998a,b; Guerra \& Daly 1996, 1998; Guerra 1997;
Daly, Guerra, \& Wan 1998; Guerra, Wan, \& Daly 1998; Wan 1998). 
For example,  detailed results on the density and temperature 
of the ambient medium of the FRII
sources are presented by WDW97a,b, and the redshift evolution of
the 
characteristic size of an FRII source and its application for
cosmology 
are presented by Guerra \& Daly (1996, 1998).

This paper presents results on several key parameters
of the FRII sources and their environments that were not
presented and discussed in previous papers.  
These parameters include
the luminosity in directed kinetic energy
of the jet, also known as the beam power, $L_j$, the total time the AGN
will produce the collimated outflow, $t_*$, 
the thermal pressure of the
ambient medium, $P_{th}$, the effect of the hot gas on the
microwave background radiation, 
and the total gravitational mass of the host cluster (we find
that the radio sources lie at the centers of gas rich clusters of
galaxies).  
The way that these parameters can be estimated
from radio data, and the empirical results for the sample are
presented here.  

The paper is structured as follows. A brief
description of the sample and data for the sample are presented in
\S\ref{sec:sample}. 
Discussions on the jet luminosity and timescale of jet activity
are given in \S\ref{sec:lj}, the thermal
pressure of the ambient medium and estimates of the effect of the
hot gas on the microwave background radiation are
presented in \S\ref{sec:pth}, and estimating the total 
gravitational mass of the
the host cluster of the radio source is discussed in
\S\ref{sec:mass}.  
Each of these sections are divided into two subsections, with the
first subsection presenting theory and the second one presenting empirical
results for the sample. Further discussions of the results and a
summary of the paper are presented in \S\ref{sec:sum}.

Values of Hubble's constant 
$H_o=100~h~\hbox{km s}^{-1} \hbox{Mpc}^{-1}$ and the de/acceleration
parameter $q_{o}$=0 (an open empty universe with zero
cosmological constant) are used to estimate all the parameters in
this study.  Different choices of cosmological
parameters, within reasonable limits, 
give results that are very similar to, and consistent
with, those presented here.  

\section{\label{sec:sample}Sample Description and Data}

The sample used in this study has been described in detail in
several previous papers (\eg WDW97a,b; Guerra \& Daly 1998). 
The readers are referred to these papers for a full description of the sample. 
A brief description of the sample is as follows. All the FRII sources
in this sample are  very powerful 3CR sources, having 178 MHz radio
power greater than $3\times 10^{26} \hbox{ } h^{-2} \hbox{ W
Hz}^{-1} \hbox{sr}^{-1}$. They are drawn from the samples of 
Leahy, Muxlow, \& Stephens (1989; hereafter LMS89), which
contains sources with large angular sizes, and
Liu, Pooley, \& Riley (1992; hereafter LPR92), which contains
sources with smaller angular sizes. The final sample contains 27
radio lobes from 14 FRII radio galaxies, and 14 lobes from 8 FRII
radio loud quasars.  These sources have redshifts ranging from zero
to 2, and projected core-hot spot separations between $25 h^{-1}$ kpc and 
$200 h^{-1}$ kpc.   Note that the questions of projection effects
and radio power selection effects have been addressed in great
detail by Wan \& Daly (1998a,b).

WDW97a,b used the radio maps from LMS89 and LPR92 
to estimate the width of the radio
bridge behind the radio hot spot, $a_L$, the non-thermal pressure
of the radio lobe that drives the forward shock front, $P_L$,
and the rate at which the bridge is lengthening, referred to as
the lobe propagation velocity $v_L$.  The radio
information may also  be used to estimate the beam power and total time
the source will be active, as described in \S 3.  

In order to estimate the Mach number of lobe advance, which in
turn may be combined with the lobe propagation velocity $v_L$ to
estimate the temperature of the ambient gas, high
quality maps which image large portions of the radio bridge
are required; the radio
bridge is defined as the radio emitting region that lies between
the radio hot spot and the origin of the host galaxy or quasar.  
There are 16 bridges in the
sample for which we have a detection of the Mach number of lobe
advance, which includes 13 radio galaxy bridges and 3 radio loud quasar
bridges.  Thus, the ambient gas temperature and pressure may be
estimated for these sources.  
Lower bounds on the Mach number, and hence upper
bounds on the ambient gas temperature and pressure,
are available for the 14 other bridges, including 8 galaxy bridges and 6 quasar
bridges, as described in WDW97a,b.  
The maps of the sources do not have sufficient dynamic range
to allow detailed analyses of their
bridge structure, and thus do not have estimates on the Mach
number of lobe advance, nor on the ambient gas pressure and
cluster mass.  
As a result, the sample of sources
with estimates of the ambient gas pressure and cluster mass is
rather small.  

Results for radio-loud quasars and radio galaxies are analyzed separately 
when there are enough data points, as is the case in the study of
the beam power. As noted above,
the number of sources with thermal pressure and cluster mass estimates 
is rather small (16 lobes with detections  and 14 with bounds).
Thus it is not practical to separate quasars and galaxies in the
study of these parameters.
 
Cygnus~A (3C 405) is the only source with a redshift $\sim 0$ 
in our sample, and in some cases, results 
seem to depend on whether or not this source is included.
Thus, subsamples both with and 
without Cygnus A are analyzed.

\section{\label{sec:lj}Luminosity in Directed Kinetic Energy of the Jet}

\subsection{\label{ssec:lj-t}Theory}

The jet or beam of an FRII radio source refers to the collimated outflow
which carries energy from the AGN.   
When the jet impinges upon the external medium, a strong shock is formed and 
the kinetic energy of the jet is deposited in the vicinity of the
shock front, which is marked by the radio hot spot. In principle,
one can calculate the beam power, $dE/dt$, carried by the jet by studying
the propagation of the hot spots. However, in practice it is difficult to use 
hot spot properties to estimate the beam power,  since
the hot spot is generally not resolved, and 
the position and properties of radio hot spots vary on short time scales 
(Laing 1989; Carilli, Perley, and Dreher 1988; Black et al.
1992.).

The model given by (Daly 1994, 1995) bypasses these difficulties
by using properties of the more stable radio lobes to estimated
the luminosity in directed kinetic energy, or the beam power, $L_j$.
Observations and numerical simulations indicate that radio lobes of
powerful extended radio sources propagate
supersonically relative to the ambient medium 
(Alexander \& Leahy 1987; Prestage \& Peacock 1988; 
Cox, Gull, \& Scheuer 1991; Daly 1994; WDW97a,b). 
Leahy (1990) shows that the properties of the radio lobe can be
used to estimate the rate of energy input. His equation may be
rewritten as:
\begineq
	L_j = 4\pi a_L^2 v_L P_L.
\label{eq:lj1}
\endeq
Here $a_L$ is the cross-sectional radius of the radio lobe,
$v_L$ is the lobe propagation velocity, and the
FRII source is assumed to be cylindrically symmetric about the radio axis.
Using typical units, $L_j$ can be expressed as
\begineq
        L_j = 3.6 \times 10^{44} \mbox{erg s}^{-1}
          \left( {a_L \over \mbox{kpc} }\right)^2
          \left( {v_L \over c}\right)
          \left( {P_L \over 10^{-10}\, \mbox{dyne cm}^{-2}} \right),
\label{eq:lj2}
\endeq

All three parameters used to estimate $L_j$ in eq.~(\ref{eq:lj2})
can be estimated from radio data (see, for example, Daly 1995, or
WDW97a,b).   
The lobe radius $a_L$
can be readily measured from the radio map (see WDW97a,b for
detail). The lobe propagation velocity $v_L$ can be 
estimated using multi-frequency observations of radio lobes
and bridges, by modeling the spectral aging of relativistic electrons due to
synchrotron and inverse Compton losses (\cf WDW97a,b; Wan \& Daly
1998a,b; Myers \& Spangler 1985; Alexander \& Leahy 1987; LMS89;
LPR92; Carilli \etal 1991).   
The lobe pressure $P_L$ is given by
\begineq
	P_L = \left( \frac{4}{3} b^{-1.5} + b^2 \right)
		{ B_{min}^2 \over 24 \pi},
\label{eq:pl}
\endeq
where $B_{min}$ is the magnetic field strength in the radio lobe
under the minimum-energy condition (\cf Miley 1980; Pacholczyk 1970), 
$b$ is the ratio of the true magnetic field strength $B$ to 
the minimum-energy magnetic field strength: $B = b B_{min}$, and
the equation is in cgs units. 
WDW97b estimate $b$ to be about
0.25, with a source-to-source dispersion less than about $15\%$. This
value of $b$ is consistent with values obtained by other authors
(\cf Carilli \etal 1991; Perley \& Taylor 1991; Feigelson \etal
1995), and is used throughout this study.

The radio spectral index of an FRII source is used in the spectral
aging analysis to estimated the lobe propagation velocity $v_L$.
It is shown in WDW97b that there is a clear correlation between
the radio spectral indices of the sources in this sample and
redshift.
The spectral index $\alpha$ can be expressed as
$\alpha \propto (1+z)^s$. The best fitted value of $s$ is
$0.8\pm0.2$ with a reduced $\chi^2$ of 1.7.  Such a correlation could be
caused by intrinsic curvature in the initial electron energy
spectrum or effects of inverse Compton cooling on the hot spot
spectral index (see WDW97b). Thus, the observed
spectral index may not be the appropriate index to used
in spectral aging analysis. This lead WDW97b to also
consider correcting the observed spectral indices to zero
redshift using the empirical correlation between $\alpha$ and
$(1+z)$ for the
sample. 

The beam power $L_j$ can be used to estimate the total time that
the AGN will be active, and will 
produce highly collimated outflows. Following Daly
(1994, 1995) and Guerra \& Daly (1996, 1998), the total lifetime 
of an outflow,
defined as $t_{\star}$, is related to the energy extraction rate
$L_j$ by a power law:
\begineq
  t_{\star} = C L_j^{-\beta_{\star}/3},
\label{eq:tstar}
\endeq
where $\beta_{\star}$ is a parameter that has implications for models of
energy extraction from the central engine.  
The current best
estimate of $\beta_{\star}$ is $1.75\pm 0.25$ (Guerra, Daly, \&
Wan 1998).  
(Note that
the $\beta_{\star}$ used here is not related to 
$\beta_0$ used in \S\ref{sec:pth} and \S\ref{sec:mass}
for the King density profile.) The characteristic size of an FRII
source is related to its lifetime as $D_{\star}=v_L t_{\star}=C
v_L L_j^{-\beta_*/3}$, where $C$ is
the normalization in eq.~(\ref{eq:tstar}). This normalization
factor is chosen so 
that at $z \approx 0$, the characteristic size of Cygnus A 
matches the observed average lobe-lobe size for
sources at this redshift. That is, $(v_L)CL_j^{-\beta_*/3}$ for
one side of Cygnus A is added to $(v_L)CL_j^{-\beta_*/3}$ on the
other side of Cygnus~A, and this is set equal to $2(<D>_{z=0})$ =
$2 \times (68 \pm 14)$ kpc.
This equation is then solved for the normalization factor $C$;
the uncertainty on $C$ is indicated in Table 1.  
This normalization is then used in equation 
(\ref{eq:tstar}) to obtain an estimate of the total time 
that the AGN will be producing large-scale jets.  The value for
$t_*$ estimated using independent information from each side of
the source should of course be equal, and are generally very
close in value.  
The value adopted for the beam power, $L_j$, of Cygnus A used above 
is listed in Table 1, $v_L$ for Cygnus A is obtained
from Table 1 from WDW97b, and a value of 
$<D>_{z=0}$ = $(68 \pm 14)$ kpc 
is adopted from Table 3 of 
Guerra, Daly, \& Wan (1998).

\subsection{\label{ssec:lj-r}Empirical Results}

The beam power, $L_j$, has been estimated using
eq.~(\ref{eq:lj2}) for the 
lobe propagation velocity estimated using the 
observed radio spectral index, and the 
redshift-corrected radio spectral index; are both listed in
Table~\ref{tab:data}. Typical values of $L_j$ are 
$10^{45}\, \mbox{$h^{-2}$ erg s}^{-1}$. 

No strong correlation is found between the beam power $L_j$ and
the linear size of the source, represented by the core-hot spot
separation $r$. 
Figure~\ref{fig:ljr} is a log-log plot of $L_j$ vs. $r$, where
the observed radio spectral index $\alpha$ is used. 
The figure of $L_j$ vs. $r$ after applying the
redshift-correction on $\alpha$ is almost identical to Figure~\ref{fig:ljr},
and is thus not shown here. 

For radio-loud quasars,  the best-fitted line 
in Fig.~\ref{fig:ljr} has a slope of about zero. 
A Spearman rank correlation analysis shows that the correlation
coefficient between $L_j$ and $r$ is $-0.03 (7.72\%)$ when no
redshift-correction on the spectral index is applied, and $-0.01
(1.78\%)$ when the spectral index correction is applied. Here the
number in parenthesis is the significance of the correlation.
These results suggest that the correlation 
between $L_j$ and $r$ for radio-loud quasars is insignificant, 
either with or without the
$\alpha\!-\!z$ correction. That is, $L_j$ is 
independent of $r$ for the radio-loud quasars. 

For the radio galaxies, $L_j$ appears to increase slightly with $r$ in
Fig.~\ref{fig:ljr}, and results from the Spearman rank correlation
analysis hint a marginally significant correlation between $L_j$
and $r$. The correlation coefficient is about 0.3, with a
significance level of about $90\%$, which holds with or without
the $\alpha\!-\!z$ correction. 
However, note that the best-fitted line for radio galaxies in
Fig.~\ref{fig:ljr} has a slope only $2\sigma$ away from zero, and
that a correlation coefficient of about 0.3 is rather
weak. Thus it appears that $L_j$ is at most weakly dependent 
on $r$ for the radio galaxies. 
These results for radio-loud quasars and radio galaxies are consistent with 
the assumption that
$L_j$ is roughly constant over a source's lifetime.

The relationship between $L_j$ and redshift and radio power
($P_r$) has been studied in detail in Wan \& Daly (1998a). 
Increases of $L_j$ with $z$ and with $P_r$ are observed (see
Figures~\ref{fig:ljz} and \ref{fig:ljp}). However, the
radio power of the sources in the sample also increases with
redshift since the sources used in this study 
all come from the flux-limited 3CR survey. 
As a result, which of the $L_j-z$  and $L_j-P_r$ 
correlations is more significant, and the role of 
radio power selection effects, needs to
be considered carefully. 
Wan \& Daly (1998a) used two parameter fitting and partial rank
analysis to investigate this.
The two parameter fit 
\hbox{$L_j \propto P_{178}^{n_p} (1+z)^{n_z}$} yields the
following $n_z$ values: $n_z=1.45\pm0.32(0.60)$
 for all galaxies, $n_z=3.83\pm0.93(1.73)$ for all galaxies except
for Cygnus A, and $n_z=9.56\pm1.37(3.91)$ for all radio-loud
quasars. 

The number in parenthesis is the uncertainty on
$n_z$ times $\sqrt{r\chi^2}$, which includes the effect of a
reduced $\chi^2$ $(r \chi^2)$ that is greater than one for the fit. This
notation will be used throughout this paper.

The values of $n_z$ are
more that $2\sigma$ away from zero, for both radio galaxies
and radio-loud quasars. This means that the $L_j-z$ correlation
is unlikely to be caused purely by radio power selection effects. That is, 
a real increase of $L_j$ with redshift exists, though the
magnitude of this redshift evolution is not well determined.

The $L_j-P_r$ correlation, on the other hand, is rather poorly constrained. 
Results from the one parameter fit suggest $L_j$ goes
roughly proportional to $P_{178}$ (Figure~\ref{fig:ljp}).
The two parameter fit 
\hbox{$L_j \propto P_{178}^{n_p} (1+z)^{n_z}$}, which takes into
account the redshift dependence of radio power, gives values of
$n_p$ with large uncertainties: $n_p=0.76\pm0.11(0.20)$
 for all galaxies, $n_p=0.25\pm0.22(0.41)$ for all galaxies except
for Cygnus A, and $n_p=-0.54\pm0.29(0.82)$ for all radio-loud
quasars.
Thus, the exact relationship between $L_j$ and radio power is not clear
at the present time.
Note, however, large amounts of scatter are seen in the $L_j$
vs. $P_{178}$ plot (Figure~\ref{fig:ljp}), which is also reflected
in the large reduced $\chi^2$ of the one parameter fit. 
For the same beam power, the radio power can vary by about a factor of ten.  
This suggests that radio power
is not an accurate measure of the beam power.

The fact that $L_j$ increases with redshift implies that the
total lifetime of an outflow, $t_{\star}$, estimated using
eq.~(\ref{eq:tstar}), decreases with redshift. The values of
the full lifetime of the source 
$t_{\star}$ are listed in
Table~\ref{tab:data}. Typical values of $t_{\star}$ are about $10^7$
years.  Figure~\ref{fig:tstarz} plots $t_{\star}$  as a function of
redshift. It is clear that $t_{\star}$ decreases with $z$, which
can explain the fact that the average size of powerful extended
3CR sources decreases with redshift for $z>0.3$ (see Guerra \&
Daly 1998), while the lobe propagation velocity $v_L$ increases
with redshift. Clearly if $v_L$ increases with redshift, and the 
mean source size $D_*$ decreases with redshift, then $t_*$ must
decrease with redshift since $D_* \simeq v_L t_*$. 
The sources at high redshift produces more powerful jets (with
larger $L_j$) for a shorter period of time (with smaller $t_*$), 
which results in smaller average
sizes than low-redshift sources.

\section{\label{sec:pth}Thermal Pressure of the Ambient Gas}

\subsection{\label{ssec:pth-t}Theory}

The studies of WDW97a and b show that both the ambient gas
density and temperature of an FRII source can be estimated from radio data.
The thermal pressure of the ambient gas $P_{th}$ 
is obviously proportional to the
product of the density $n_a$ and the temperature $T$.  Interestingly, as
shown below, although the lobe propagation velocity enters into
$n_a$ and $T$, it cancels out in the product, so the thermal
pressure of the ambient gas can be estimated using single
frequency radio data if the radio emission from the bridge is
mapped over a large enough region of the bridge.  
This offers a completely new method to estimate the thermal 
pressure of the ambient gas surrounding high-redshift radio
sources using only single frequency radio data.  

It can be complicated and time consuming to estimate the ambient gas density
in the vicinity of classical double radio sources using X-ray observations,
especially for sources at high redshift. An attractive
alternative method of estimating $n_a$ in the vicinity of a
radio source is to use the radio data.  
Given that the radio lobe represents a strong shock front, the
lobe pressure $P_L$, the ambient gas density $\rho_a$, and
the lobe propagation velocity are related: 
the strong shock jump conditions imply that 
$P_L \approx 0.75\rho_a v_L^2 $ holds, where $P_L$ is 
the non-thermal pressure inside the radio lobe,
and $\rho_a$ is the mass density of the ambient gas. 
The electron number density of the ambient gas $n_a$ 
can be estimated using
\begineq
	n_a = {P_L \over 1.4 \mu m_p v_L^2},
\label{eq:na}
\endeq
where $\mu$ is the mean molecular weight of the gas in AMU, with a value of
0.63 when solar abundances are assumed, and $m_p$ is the proton
mass.

The ambient gas temperature can be estimated by careful studies of
the radio bridge, as discussed in detail in WDW97a. A brief
description of the theory is as follows.
When the non-thermal pressure of the radio bridge is 
much greater than the ambient gas pressure, 
the bridge undergoes supersonic expansion. This lateral
expansion can be treated as a blast wave and the expansion
velocity is determined by ram pressure confinement. This causes the width of
the bridge to vary with time approximately as $t^{1/2}$
(\eg Begelman \& Cioffi 1989; Daly 1990; 
WDW97a). Since the lobe propagation velocity $v_L$ is 
roughly constant during a given 
source's lifetime (\eg LPR92; WDW97a,b), the lobe front will
advance a distance of $v_L t$ during $t$ as the bridge expands.
This means that at a distance $x$ from the hot spot, the bridge
expansion time is $t=x/v_L$. Thus
the shape of the radio bridge roughly follows a square root law, with
the bridge width $a(x) \propto x^{1/2}$. This behavior was in
fact found by WDW97a.  
As the bridge expands, the 
pressure inside decreases until it becomes comparable to the
ambient gas pressure. When this occurs, the lateral expansion velocity
becomes sonic, causing a break in the functional form of $a(x)$.  
The width of the bridge starts to deviate from the square root law and 
becomes roughly constant as equilibrium is being reached. 
At the point where such a break occurs, the lateral expansion velocity is
approximately sonic. Thus the sound speed of 
the ambient gas ($c_s$) can be estimated using 
\begineq
	c_s \approx \left. v_s \right \vert_b 
		      = \left.{da \over dt}\right\vert_b 
		      = \left.{da \over dx}\right\vert_b {dx\over dt} 
		      = \left.{da \over dx}\right\vert_b \cdot v_L ,
\label{eq:cs}
\endeq
where $\left. v_s \right\vert_b$ denotes the lateral expansion 
velocity at the break.

The Mach number of lobe propagation is defined as $M \equiv v_L/c_s$. 
Given the expression for
$c_s$ in eq.~(\ref{eq:cs}), the Mach number of lobe propagation is:
\begineq
	M \equiv {v_L \over c_s} 
	  \approx \left( \left.{da \over dx}\right\vert_b
		\right) ^{-1} = {2 x_b \over a_b},
 \label{eq:mach}
\endeq
where $a_b$ is the width of the bridge at the break, and $x_b$ is
the position of the break relative to the hot spot (see WDW97a).  
It can be seen from eq.~(\ref{eq:mach}) that $M$ depends only on 
the geometrical shape of the radio bridge. 

The ambient gas temperature T can be expressed as:
\begineq
        T = {\mu m_p\over \gamma k} c_s^2
        \propto
         \left (\left.{da \over d x} \right\vert_b \right)^2 \cdot v_L^2
         \propto v_L^2 M^{-2},
\label{eq:t}
\endeq
where $\gamma$ is the ratio of specific heats of the ambient gas.
                                                                  
Multi-frequency radio maps are required in order to use
eqs.~(\ref{eq:na}) and (\ref{eq:t}) to estimate 
the ambient gas density and temperature of an FRII source,
since both $n_a$ and $T$ depend on $v_L$. However, there is no such
requirement in order to estimate 
the  thermal pressure of the ambient gas. 
The ambient gas pressure in the vicinity of the radio lobe is
simply
\begineq
	P_{th} = (n_a+n_i) k T
	       \propto \left( {P_L \over v_L^2} \right) v_L^2
			M^{-2} 
	       \propto {P_L \over M^2},
\label{eq:pth1}
\endeq	
with $P_L$ estimated using eq.~(\ref{eq:pl}) and $M$ estimated using 
eq.~(\ref{eq:mach}). Here $n_i$ is the number densities of ions,
and for a gas with solar abundances, the electron density is
$n_a=1.21 n_i$.
Substituting in the numerical constants in eqs.~(\ref{eq:na}) and
(\ref{eq:t}),  the thermal pressure is given by
\begineq
 	 P_{th}=0.8 {P_L \over M^2},
 \label{eq:pth2}
\endeq
for a gas with a specific heat ratio $\gamma=5/3$.

The thermal pressure estimated using eq.~(\ref{eq:pth2}) 
depends only on the lobe pressure $P_L$ and the geometrically determined Mach
number $M$, both of which can be obtained from 
single frequency observations of the radio lobe and
bridge.
No spectral aging analysis is needed in order to estimate
$P_{th}$, since the dependences of $n_a$ and $T$ on $v_L$ cancel.
As a result, $P_{th}$ is also independent of whether or
not the redshift-correction on the radio spectral index is
applied.  

At low-redshift, X-ray data can often be used to estimate $n_a$ and
$T$, and hence $P_{th}$. 
However, high-resolution X-ray data are often difficult to obtain 
for sources at high redshift. 
This new method of estimating $P_{th}$, using single frequency radio
data, offers an attractive alternative.
It provides a powerful tool to study the environments
of powerful classical double radio sources, 
especially those at high-redshift.  Since it has been shown that
these sources are in cluster-like gaseous environments, the
sources may be used to study evolution of gas in clusters of 
galaxies (as discussed by Daly 1995; WDW97a,b).  

\subsection{\label{ssec:pth-r}Empirical Results}

The value of $P_{th}$, the thermal pressure of the ambient gas in the
vicinity of the radio lobe, is listed in Table~\ref{tab:data}.
Most sources in the sample have $P_{th}$ on the order of ($10^{-11}$ to
\hbox{$10^{-10})\,h^{4/7}\,\mbox{dyne cm}^{-2}$}, 
which is typical of  gas in low-redshift clusters of
galaxies.  
This is consistent with results obtained by WDW97(a,b),
who find cluster-like density and temperature for 
the ambient gas of the FRII sources
in the sample. Note that the thermal pressures obtained here do
not depend on a spectral aging analysis, whereas 
density and temperature estimates do.

The X-cluster around Cygnus has been observed by numerous
authors (Arnaud et. al. 1984; Carilli, Perley, \& Harris 1994).
The thermal pressure of the ambient gas near its radio lobe is
estimated to be about \hbox{$10^{-10}\, \mbox{dyne cm}^{-2}$}
(see Carilli, Perley, \& Harris 1994). 
This estimate from X-ray data is consistent with the thermal
pressure estimated here for Cygnus A using the new method.

Figure~\ref{fig:pthr} plots $P_{th}$ as a function of 
the core-hot spot separation $r$. 
It can be seen that $P_{th}$ decreases with $r$. 
A negative pressure gradient is expected  for 
an isothermal gas distribution that follows the
King density profile,  which is suggested to be
the ambient gas distribution for sources in our sample (WDW97a,b). 
For this gas distribution,
the thermal pressure decreases with $r$ as
\begineq
 	P_{th}(r)=
	 P_{th,\,c} \left [1+\left ( {r \over r_c}\right
 	   )^2\right]^{-{3 \over 2} \beta_0},
\label{eq:pthr}
\endeq
where $P_{th,\,c}$ is the thermal pressure at the center of the
cluster, and $r_c$ is cluster core radius.

The pressure gradient seen Figure~\ref{fig:pthr} appears to be
consistent with
that expected for an cluster gas distribution that is isothermal
and follows a King density profile, with a cluster core radius $r_c$ of 
\hbox{$\sim (50 \mbox{ to } 150)\, h^{-1}$ kpc}.
The sources in our sample have values of $r$ ranging from 
\hbox{(25 to 250) $h^{-1}$ kpc}.  
Within this radius range, the pressure profile
given by eq.~(\ref{eq:pthr}) has an average slope of 
\hbox{($\sim\! -1.4 \mbox{ to } -0.7$)}, 
for $r_c$ values of \hbox{(50 to $150\,h^{-1}$ kpc)},
assuming $\beta_0=2/3$. These expected
slopes are consistent with the best-fit slopes of 
\hbox{$\sim \!-1.0\pm0.3$} in Figure~\ref{fig:pthr}.

There is some hint of a redshift evolution of
the cluster core radius from the data. Figure~\ref{fig:pthrhilo} plots $P_{th}$
vs. $r$ in two redshift bins, $z\!<\!0.9$ and $z\!>\!0.9$. The division
at $z=0.9$ is chosen so that the two redshift bins cover about the
same range in redshift and contain about the same number of data
points.  
It appears that $P_{th}$ decreases less rapidly with $r$ at high
redshift than at low redshift, suggesting a larger $r_c$ at high
redshift. The best-fit slope of $\sim\!-1.4$ in the low-redshift
bin is consistent with \hbox{$r_c\sim 50\, h^{-1}$ kpc} 
for the pressure profile given by eq.~(\ref{eq:pthr}), 
whereas the best-fit slope of about $-0.4$ in the high-redshift
bin is consistent with \hbox{$r_c \sim 250\,h^{-1}$ kpc}.
This result is still preliminary since only sources with
detections of $P_{th}$ are included in the fits. Better estimates
of $P_{th}$ for sources currently with only upper bounds on
$P_{th}$ will help to better determine whether $r_c$ is evolving
with redshift.

The results obtained above 
are consistent with results obtained by
WDW97b. They study the ambient gas density profile and
find that $r_c\sim50\, h^{-1}$ kpc when high- and low-redshift source are
considered together. They also find that that data are consistent with
a constant core gas mass model where $r_c$ increases with
redshift roughly as $r_c\propto (1+z)^{1.6}$.  
Note that results from
the study of $P_{th}$ do not depend on spectral aging analysis, whereas those
from the study of $n_a$ do. It is thus encouraging that
consistent results are obtained from the two studies.

For the following analysis, 
we use both a non-evolving core radius of \hbox{$50\, h^{-1}$
kpc}, and an evolving core radius of \hbox{$50\,(1+z)^{1.6} h^{-1}$
kpc}, with a focus on the latter.

The thermal pressure at the center of the cluster 
($P_{th,\,c}$) can be estimated by scaling the pressure in the
vicinity of the radio lobe ($P_{th}$) to the cluster center.
This follows because the studies of Daly (1995) and WDW97a,b
indicate that these radio sources are located at the centers of
clusters of galaxies, so the core-hot spot separation can be used
as an estimate of distance from the cluster center.  
The current data on $T$, $n_a$, and $P_{th}$ are all consistent
with the gas distribution being isothermal and following a King
profile. Thus we use the pressure profile for such a gas
distribution, as given by eq.~(\ref{eq:pthr}), to estimate the
central thermal pressure.  
The central pressure is simply
\begineq
 	P_{th,\,c}=
	 P_{th}(r) \left [1+\left ( {r \over r_c}\right
	  )^2\right]^{{3\over2} \beta_0},
\label{eq:pthc}
\endeq
The values of $P_{th,\,c}$ for the sources in our sample
are listed in Table~\ref{tab:data}, where a value of $\beta_0 =
2/3$ has been assumed.   
Most sources have values of
$P_{th,\,c}$ around \hbox{$10^{-10}\,h^{4/7}\, \mbox{dyne cm}^{-2}$}, 
which is rather typical of gas pressure in the core regions of 
low-redshift galaxies clusters.

Results on the redshift evolution of the central gas pressure
$P_{th,\,c}$ are somewhat uncertain because $P_{th,\,c}$ seems to
correlate with both redshift and radio power when either
correlation is considered separately (see Figures~\ref{fig:pthcz}
and \ref{fig:pthcp}), and it is not clear
at present which correlation is more significant.  
For the non-evolving core radius model, a two-parameter fit 
of $P_{th,\,c} \propto  (1+z)^{n_z} P_{178}^{n_p}$
yields $n_z=0.29 \pm 0.49 (0.67)$ and $n_p=0.66\pm0.19 (0.26)$ when
all sources are included, and $n_z=1.51\pm2.11 (2.99)$
and $n_p=0.34\pm0.58 (0.82)$ when Cygnus~A is not included.
For the evolving core radius model, the two-parameter fit 
gives $n_z=-1.22\pm0.49 (0.65)$ and $n_p=0.88\pm0.19 (0.25)$ when
all sources are included, and $n_z=1.56\pm2.11 (2.82)$
and $n_p=0.15\pm0.58 (0.78)$ when Cygnus~A is not included.
Note that Cygnus~A appears to lie in a cooling-flow region
while most of the other sources are
not in cooling-flow regions (see WDW97a; \S\ref{sec:mass}).
Thus fits with and without Cygnus~A are both performed.  
In any case, the large uncertainties on $n_z$ and $n_p$ make it
hard to determine the magnitude of the redshift evolution of
$P_{th,\,c}$.

The thermal pressure can be used to predict the amount of cosmic
microwave background (CMB) diminution, also known as
the Sunyaev-Zel'dovich (S-Z) effect,  that is expected from the cluster. 
The reduction in the cosmic radiation ($\Delta I_{\nu}$) in the direction of a
cluster is given by (\eg Sunyaev \& Zel'dovich
1980; Sarazin 1988; Rephaeli 1995) 
\begineq
  \Delta I_{\nu}/I_{\nu} = G(\omega) y.
\label{eq:dinu1}
\endeq
Here $I_{\nu}$ is the specific intensity of the CMB at the
observing frequency $\nu$, $y$ is the Comptonization parameter,
and the function $G(\omega)$ is 
\begineq
	G(\omega)={\omega e^{\omega} \over e^{\omega}-1} \left [
 	\omega \left ({e^{\omega}+1 \over e^{\omega}-1}\right )-4 \right],
\label{eq:gomega}
\endeq
where $\omega\equiv h\nu/kT_r$, and $T_r=2.73\,\mbox{K}$ is the
CMB temperature (Mather \etal 1990).

The Compton $y$ parameter is given by
\begineq
   y =\int {kT \over m_e c^2} \sigma_T n_e dl
     ={\sigma_T \over m_e c^2}
	\left ( {n_e \over n_e + n_i} \right ) \int P_{th} dl.
\label{eq:ycom}
\endeq
where $\sigma_T=(8\pi/3)[e^2/(m_e c^2)]^2$  is the Thompson
electron scattering cross section, $T$ is the cluster gas
temperature, $n_e$ is the electron density,
and $n_i$ is the ion density.
Using the thermal pressure in a cluster with 
an isothermal-King gas distribution (eq.~[\ref{eq:pthr}]), 
this gives 
\beginmlet
\begineq
	y \approx 6.3\times 10^{-5} 
          \left ( {\mbox{\raisebox{4pt}{$P_{th,\,c}$} } \over 
	  	\mbox{\raisebox{-4pt}{$10^{-10}$\,dyne cm$^{-2}$} } }\right )
	  \left ( {\mbox{\raisebox{4pt}{$r_c$ }} \over 
		\mbox{\raisebox{-4pt}{0.25\,Mpc}} } \right ) 
	{\mbox{\raisebox{4pt}{$\Gamma(3\beta_0/2-1/2)$ }} \over 
		\mbox{\raisebox{-4pt}{$\Gamma(3\beta_0/2)$ }} }
	  \, (1+x^2)^{-3\beta_0/2+1/2},
\label{eq:ycom1} 
\endeq
where $x \equiv r/r_c$, and the gas is taken to have solar
abundance so that $n_e=1.21n_i$.  For a typical value of
$\beta_0=2/3$, 
\begineq
	y  \approx 1.12\times 10^{-4} 
          \left ( {P_{th,\,c} \over 10^{-10}\, \mbox{dyne cm}^{-2}} \right )
	  \left ( {r_c \over 0.25 \, \mbox{Mpc}} \right )
	  (1+x^2)^{-1/2}.
\label{eq:ycom2} 
\endeq
\endmlet

At low frequency, clusters with strong radio sources are
generally avoided for measurements of the Sunyaev-Zel'dovich
effect because emission from the radio source masks the microwave
diminution.
Such contaminations from radio sources are reduced at high frequency
since emission from steep-spectrum radio sources,
such as the FRII sources studied here,
decreases rapidly with increasing frequency.
 %
The Sunyaev-Zel'dovich Infrared Experiment (SuZIE) can measure 
the S-Z effect around 140 GHz, where the amount of CMB
intensity diminution is near its peak. 
At this frequency,  a cluster with an isothermal-King density profile
will cause a CMB  intensity diminution of about 
\begineq
	\begin{array}{ll}
	\Delta I_{\nu} \approx  &
         7.0\times 10^{-19} 
	  \mbox{erg s$^{-1}$Hz$^{-1}$cm$^{-2}$sr$^{-1}$ }
          \left ( {\mbox{\raisebox{4pt}{$P_{th,\,c}$} } \over 
	  	\mbox{\raisebox{-4pt}{$10^{-10}$\,dyne cm$^{-2}$} } }\right )
	  \left ( {\mbox{\raisebox{4pt}{$r_c$ }} \over 
		\mbox{\raisebox{-4pt}{0.25\,Mpc}} } \right ) \\
	 & \times {\mbox{\raisebox{4pt}{$\Gamma(3\beta_0/2-1/2)$ }} \over 
		\mbox{\raisebox{-4pt}{$\Gamma(3\beta_0/2)$ }} }
	  \, (1+x^2)^{-3\beta_0/2+1/2},
	\end{array}
\label{eq:sz140g1} 
\endeq
and for a typical value of $\beta_0=2/3$, 
\begineq
	\Delta I_{\nu} 
         \approx 1.24\times 10^{-18} 
	  \mbox{erg s$^{-1}$Hz$^{-1}$cm$^{-2}$sr$^{-1}$ }
          \left ( {P_{th,\,c} \over 10^{-10}\, \mbox{dyne cm}^{-2}} \right )
	  \left ( {r_c \over 0.25 \, \mbox{Mpc}} \right )
	  (1+x^2)^{-1/2}.
\label{eq:sz140g2} 
\endeq

For a detector with a FWHM beam size $\theta_b$, the total amount of
CMB diminution within the beam, defined as $\Delta
F_{\nu}$, is roughly
\begineq
	\Delta F_{\nu} 
         \approx f \cdot 24\, \mbox{mJy} 
	 \left ( {\theta_b \over \mbox{1.7 arcmin}} \right )^2
	  \left ( {P_{th,\,c} \over 10^{-10}\, \mbox{dyne cm}^{-2}} \right )
	  \left ( {r_c \over 0.25 \, \mbox{Mpc}} \right ),
\label{eq:sz140g3} 
\endeq
where $f$ is the beam dilution factor, defined
as the ratio of the average $\Delta I_{\nu}$ within the
beam to the peak value at the cluster center; and
the SuZIE FWHM beam size at 140 GHz of 1.7 arcmin (\eg Holzapfel \etal
1997) is used to calculate the numerical value. 
For a Gaussian beam,  when the HWHM beam size corresponds to 1$r_c$,
$f\approx0.7$, and when the HWHM beam size corresponds to 2$r_c$, 
$f \approx 0.5$ (see Rephaeli 1987).   
At $z<2$, the SuZIE beam radius is less than $\sim 320\, h^{-1}
\mbox{kpc}$ for $q_0=0$ with no cosmological constant.
Thus a cluster with $P_{th,\,c}\!\sim\! 10^{-10}\, \mbox{dyne
cm}^{-2}$ and $r_c\! \sim \,0.25\,\mbox{Mpc}$ will have a $\Delta F_{\nu}$
of about (15 to 20) mJy within the SuZIE beam. 
The clusters surrounding the FRII sources
in our sample have $P_{th,\,c}$
on the order of $10^{-10}\,h^{4/7}\,\mbox{dyne cm}^{-2}$ and the data also
suggest that $r_c$ increases with redshift, reaching about 
\hbox{$300\, h^{-1}\,\mbox{kpc}$} at $ z\!\sim\! 2$. 
The expected CMB diminution within the SuZIE beam at 140 GHz
for these clusters ranges from several to tens of mJy.
These clusters make good candidates for SuZIE observations
if the fluxes from the radio sources at 140 GHz or their
uncertainties are small compared with the expected S-Z effect signals. 

We are currently in the process of searching for high frequency
data on the radio sources in our sample in oder to identify
possible SuZIE observation candidates.
One likely candidate that comes from a preliminary search
in the published literature is 3C239. 
The expected CMB diminution
for the cluster surrounding it is about $30\,h^{-3/7}$ mJy within
the SuZIE beam at 140 GHz. In contrast,
extrapolation of the observed spectrum of 3C239 as
given by observations at 178 MHz, 10.7 GHz, and 14.9 GHz
(Kellermann \& Pauliny-Toth 1973; Genzel \etal 1976; Laing,
Riley, and Longair 1983), gives a
140 GHz flux of only 6 mJy. 
This is likely to be an overestimate since 
synchrotron aging and inverse Compton cooling can both cause the
high frequency spectral index to become steeper than that at low
frequency. The effects of inverse Compton cooling can be especially 
important since this source is at 
high redshift ($z=1.79$), where the energy density of the microwave
background is much higher than at low redshift.
Thus it appears that emission from this
radio source is weak compared with the expected CMB diminution.
Observations of the radio source at more
frequencies above 14.9 GHz can help to better constrain its 
\hbox{140 GHz} flux.

\section{\label{sec:mass}Gravitational Mass of the Host Cluster}

\subsection{\label{ssec:mass-t}Theory}

The powerful classical double radio 
sources studied here are in cluster-like gaseous
environments (see WDW97a, b). The studies of these sources
provide information on density and temperature of the ambient
gas.  Thus it is possible to estimate 
one of the most important parameters of the cluster, the total
gravitational mass, including dark matter.

The total mass can be estimated
using the density and temperature profile of the intracluster
medium (ICM) if the gas is in hydrostatic equilibrium. The sound
crossing time in the ICM is usually short compared to the
age of a high-redshift cluster, and the morphology of the X-ray
emission from many low-redshift clusters is often smooth. Thus it is
generally believed that hydrostatic equilibrium is a good
approximation of the state of the ICM for many clusters that are 
not cooling flow clusters.  
Assuming spherical symmetry, hydrostatic equilibrium 
requires
\begineq
	{1\over \rho_g} {dP \over dr} = - {G M_t(r)\over r^2},
\endeq
where $\rho_g$ is the gas density and $M_t(r)$ is the total
gravitating mass within $r$. This means that
\begineq
	M_t(r)=-{k T(r) \over G \mu m_p} \cdot r \cdot 
	  \left[ {d \log \rho_g(r) \over d \log r}+ 
		{d \log T(r) \over d \log r} \right],
\label{eq:mr1}
\endeq
where $k$ is the Boltzman constant,
$\mu$ is the mean molecular weight of the gas in amu, $m_p$ is the
proton mass, and $T(r)$ and $\rho_g(r)$ are the temperature and density
profiles of the cluster gas.

The most commonly used hydrostatic model of the ICM is the isothermal
$\beta$-model (Cavaliere \& Fusco-Femiano, 1976,1978; Sarazin \&
Bahcall 1977), where the clusters gas is isothermal and 
the density profile follows a modified King
model, i.e., \hbox{$n_a = n_c [1+(r/r_c)^2]^{-3/2\beta_0}$}.
Here $n_c$ is the core density and $r_c$ is core radius of the
gas distribution. Using this model for the ICM, eq.~(\ref{eq:mr1})
becomes
\begineq
	M_t(r)= {3 \beta_0 k \over G \mu m_p} \cdot T \cdot r \cdot  
	 {(r/r_c)^2 \over 1+ (r/r_c)^2}.
\label{eq:mr2}
\endeq
Results from numerical simulations of cluster formation suggest that 
departures of the ICM from hydrostatic equilibrium are usually
small, and the mass estimated using the standard $\beta$-model is
rather accurate (\eg Navarro, Frenk \& White 1995; Schindler 1996;
Evrard, Metzler, \& Navarro 1996). 
Note however, significant temperature decline is observed to
occur at outer regions of clusters (\eg Markevitch \etal 1997).
Thus, we only use eq.~(\ref{eq:mr2}) to estimate 
the cluster mass within $r$ rather than extrapolating to large
radii. 

The mass estimated using eq.~(\ref{eq:mr2}) is not accurate
if the cluster is cooling at the point where the cluster
temperature is measured. Inside a cooling flow region, the temperature
measured does not indicate the gravitational potential, since
hydrostatic equilibrium conditions do not apply. Considering
cooling only by thermal bremsstrahlung, which dominates other
mechanisms at typical clusters temperature, the cooling
time of the cluster gas can be estimated using
\begineq
  t_{cool}\approx 2.1\times 10^{-2} 
    \left ( {T\over 10^7\,\mbox{K}} \right )^{1/2}
    \left ( {n_a\over \mbox{cm}^{-3}} \right)^{-1} \mbox{Gyr},
\label{eq:tcool}
\endeq
where $T$ and $n_a$ are the ambient gas temperature and electron density,
respectively (see WDW97a).  Note this equation does not depend on
the Hubble's constant, provided that the estimates of $n_a$ and $T$ 
are not dependent on the Hubble's constant.

Given the ambient gas temperature and
density estimates, the cooling time for the sources in our sample
can be estimated (see WDW97a).  A cooling time less than the
age of the universe at that redshift means that the FRII source
is in a cooling flow region.  To estimate the age of the universe
at a given redshift, an open empty universe is assumed, with a 
current age of 14 Gyr, which
corresponds to \hbox{$h\approx 0.7$} for $q_0=0$ with zero 
cosmological constant.

Most sources in our sample 
do not appear to be cooling at the position where the temperature
is measured. Thus their mass estimates are likely to be
reliable.  A few sources, including the
low-redshift source Cygnus~A,  appear to lie in cooling flow
regions. 

Figures~\ref{fig:mtr42e} 		
plots the total cluster mass, including dark matter, 
within radius $r$, $M_t(r)$, as a function of the
core-hot spot separation $r$. 	Since the sources appear to lie
at
the centers of clusters of galaxies (Daly 1995; WDW97a,b), the
core-hot spot separation is used to estimate the distance from 
the cluster center.  
The radio spectral index is redshift-corrected and the
cluster core radius is taken to be 
\hbox{$r_c=50 (1+z)^{1.6}\, h^{-1}$ kpc} (see \S\ref{ssec:pth-r}).
Clusters that are cooling at the position where the temperature
is measured are marked with pentagons. 
The plots of $M_t(r)$ vs. $r$ for other models, either
with or without an $\alpha\!-\!z$ correction and/or a core radius
evolution, are presented in Figures~\ref{fig:mtr02e}
through  \ref{fig:mtr02}. 		

The cluster mass increases with $r$ for all models.
The increase is more rapid for models with an evolving core radius
than models with a fixed core radius. This merely
reflects the fact that with an increasing $r_c$, more sources lie
within the core region, where the increase of mass with radius
is rapid, and varies roughly as $M \propto r^3$.  


To determine the redshift evolution of the cluster mass, 
a three parameter fit of 
\hbox{$M_t(r) \propto r^{n_1} (1+z)^{n_2} P_{178}^{n_3}$} is performed
since the cluster mass within $r$, $M_t(r)$, is a function of $r$,
and may also be affected by radio power selection effects.
Results are listed in Table~\ref{tab:ma3p} for all models considered.
A two parameter fit of 
\hbox{$M_t(r) \propto r^{n_1} (1+z)^{n_2}$} is also performed,
and results are listed in Table~\ref{tab:ma2p}.  
For completeness, fits including all sources, all sources except
Cygnus~A, and all sources that are not in cooling flow regions, 
are all listed in the table. Note that Cygnus~A appears to be in a cooling
flow region for all the models considered.
Results obtained excluding cooling flow clusters are probably the
correct ones to consider.

The sample of clusters with mass estimates is rather small,
especially when only non-cooling flow clusters are considered.
Thus it is not surprising to see that 
the best-fit values of $n_2$
have rather large uncertainties,  which 
makes it hard to draw definitive conclusions about the redshift evolution of
the cluster mass.  
Note though, the current data do not indicate
any negative evolution of the cluster mass out to a
redshift of about two, contrary to
what is expected in cosmological models with a high density
parameter $\Omega_m$.
This is consistent with results obtained by this group
(Daly 1994; Guerra \& Daly
1996, 1998; Guerra, Daly, \& Wan 1998; and Daly, Guerra, \& Wan
1998), 
who study the characteristic size of powerful classical double
radio sources and
find the data suggest a low value of $\Omega_m$, which is
greater than $3 \sigma$  (check) away from a universe with 
$\Omega_m = 1$ (see Guerra, Daly, \& Wan 1998).  

There are some indications from the data that the redshift evolution
of the cluster core radius, and the redshift-correction on the
radio spectral index are favored. The gas mass within $r$, defined as
$M_g(r)$,  can be estimated using
\begineq
	M_g(r) = 4\pi \rho_{c} r_c^3 [x-\mbox{tan}^{-1}(x)],
\label{eq:mg}
\endeq
where $\rho_{c}$ is the central gas density, $r_c$ is the core
radius, $x\equiv r/r_c$, and a typical value of $\beta_0=2/3$ is
used for the King density profile. Knowing the gas mass and the
total mass within $r$, the gas mass fraction within $r$ can be
estimated. Figure~\ref{fig:fg42e} plots the gas mass fraction
within $r$ vs. redshift, where the radio spectral index is
redshift-corrected and the core radius 
is taken to be \hbox{$r_c=50(1+z)^{1.6}\,h^{-1}$ kpc}.

The values of gas fraction shown in the
figure are consistent with observed values for inner regions of
many clusters (\eg Donahue 1996), whereas those for models
without a redshift-correction on the radio spectral index,
and without a core radius evolution seem to be low compared 
with observed values.

Several factors may contribute to the slow decrease of gas fraction with
redshift that is seen in Figure~\ref{fig:fg42e}. 
First, the increase of cluster core 
radius with redshift means that the gas fraction estimated for
the high-redshift clusters is over a larger fraction of the
cluster core than for the low-redshift clusters.  The gas fraction
estimated at high redshift mainly represents the gas
fraction inside the cluster core, and  
it is known that gas fraction increases with increasing
radius in clusters (\eg David, Jones, \& Forman 1995).
Second, the mass contribution from galaxies is higher at the
cluster center than at large radii (\eg Loewenstein \& Mushotzky
1996). Thus by sampling a larger fraction of the cluster core at
higher redshift, a larger fraction of the total baryon mass is
not taken into account at higher redshift. Further, 
the decrease in gas fraction with redshift could be
due to the stripping of gas from galaxies in the cluster core
over time.  Futhermore, the cluster gas becomes more
concentrated toward the cluster center as it cools, which also
causes the gas fraction in the cluster center to increase with
time.

These results on cluster mass and gas fraction obtained above are still
preliminary due to the small size of the sample. More sources
with estimates on the ambient gas temperature, and hence cluster
mass, will help to test these results.

\section{\label{sec:sum}Summary and Discussion}

Several key parameters of an FRII source and its gaseous
environment are studied in this paper. 

Direct estimates of the beam power, which measures the energy 
extraction rate from the AGN by the jet,
 are obtained. Typical 
beam powers of about $10^{45}\,h^{-2}\, \mbox{erg s}^{-1}$ are found
for the sources in the sample. No strong correlation is seen
between the beam power and the linear size of the source,
which is consistent with the beam power being roughly constant
throughout the lifetime of a source. 

There is a trend for the beam power 
to increase with redshift, which is significant even after 
excluding radio power selection effects. 
The magnitude of this redshift evolution, 
however, is not well determined by the
current data.
It is well known that the quasar luminosity function undergoes
strong evolution between $z\!\sim\!0$ and $z\!\sim \!2$, with the
high-redshift quasars being more luminous than their low-redshift
counterparts (\eg Schmidt \& Green 1983; Yee \& Ellingson 1993;
La Franca \& Cristiani 1997).  This suggests that high-redshift
AGNs are more  powerful than
low redshift ones. Thus it is perhaps not
surprising to find that the beam power, or energy per unit time 
channeled into the jet by the AGN, is also higher at high 
redshift.  

The relationship between the beam power and the
radio power is not well constrained after their correlations with
redshift are taken into account. The two parameter fit of 
\hbox{$L_j \propto (1+z)^{n_z} P_{178}^{n_p}$} yields values of
$n_p$ with large uncertainties. Thus it is not clear at present
how the beam power and the radio power are related. However, 
the large amount of
scatter seen in the beam power-radio power relation suggests
that radio power is not an accurate measure of beam power.

The beam power $L_j$ can be used to estimate the
total lifetime of an outflow produced by an AGN. Following Daly
(1994) and Guerra \& Daly (1996, 1998), 
the total time for which the outflow will occur, $t_{\star}$, 
is related to $L_j$: $t_{\star} \propto L_j^{-\beta_{\star}/3}$,
where $\beta_{\star}$  is estimated to be about $1.75\pm0.25$
(Guerra, Daly, \& Wan 1998).  
Typical lifetimes of about $10^7$ years are obtained using this relation 
for the sources 
studied here. 
The lifetime of the outflow decreases with redshift, which
explains the decrease of the average size of powerful extended
3CR sources with redshift. The sources at high redshift produces jets  
for a shorter period of time, which results in smaller average
sizes  than low-redshift sources.

A new method of estimating the thermal pressure of the ambient
gas in the vicinity of powerful classical double radio source 
using only single frequency radio data is presented. The pressure
is given by the product of the ambient gas density, estimated 
using ram pressure
confinement of the radio lobe, and the ambient gas temperature,
estimated using the Mach
number for the source and the lobe advance speed.  It turns out
that the lobe propagation velocity cancels out of the product
$n_aT$, and the ambient gas pressure 
can be estimated by studying the shape of the radio
bridge, and the non-thermal pressure of the radio lobe
(see \S\ref{sec:pth}).  Thus, 
the thermal pressure, the product of the ambient gas density and
temperature, depends only on the
non-thermal pressure in the radio lobe and the geometrically
determined Mach number of lobe advance, 
both of which can be estimated using single frequency
radio data. This new method to estimate the thermal pressure does
not require a 
spectral aging analysis, and
provides a powerful tool to probe the environments of FRII
sources. The thermal pressure estimated for Cygnus A using
this new method agrees with that obtained using 
X-ray data. 

Thermal pressures on the order of $10^{-10}\,h^{4/7}\, \mbox{dyne cm}^{-2}$, 
typical of gas in low-redshift clusters of galaxies,  are found for
the gaseous environments of the FRII sources studied here.
There are hints from the current data that
the gradient of the composite pressure profile is less steep 
at high redshift than at low redshift, which can be
explained by an increase of the cluster core radius with redshift. 
The current data are consistent with a core radius evolution from 
\hbox{$\sim\! 50\,h^{-1}\,\mbox{kpc}$} at $z\!\sim\! 0$ to
\hbox{$\sim\!250\,h^{-1}\,\mbox{kpc}$} at
$z\!\sim\! 2$, which agrees with results obtained from 
studies of the ambient gas density (WDW97b). WDW97b find that the
data can be described by a model where the core gas density
decreases and the core radius increases so that the core gas mass
remains roughly constant. 

The thermal pressures obtained here can be used to estimate the 
the amount of CMB diminution expected from the clusters surrounding the 
FRII sources in the sample.
Contaminations from the radio
sources can be reduced by observing at high frequency, such as
140 GHz, since emission from the radio sources 
decreases rapidly with increasing frequency.
The cluster surrounding a source in our sample can be detected by 
SuZIE observations at 140 GHz if the flux from the 
radio source at this frequency or its uncertainty  is small
compared with the expected S-Z effect signal. A search for
high-frequency data of the sources in our sample is ongoing
in oder to identify possible SuZIE observation candidates.
Preliminary results suggest 3C239 to be a likely candidate.

The gravitational or total mass of the surrounding cluster is
estimated for the sources in the sample, assuming hydrostatic
equilibrium conditions for the gas. Masses of up to $10^{14}
M_{\odot}$ are found for the central regions 
\hbox{($r \lesssim 250 \,h^{-1}\,\mbox{kpc}$)} of the clusters, 
consistent with typical values for low-redshift clusters.  
The redshift evolution of the cluster mass is not well
determined.  Current data do not indicate any negative
evolution of the cluster mass, contrary to what is expected in a
high density universe. This is consistent with results obtained
by members of this group 
(Guerra \& Daly 1996, 1998; Guerra, Daly, \& Wan 1998; Daly,
Guerra, \& Wan 1998), who study the characteristic size of 
FRII sources and find the data strongly favor a low  value for
the density parameter
$\Omega_m$, which is greater than $3 \sigma$ below unity. 
Note however,
we only study the central regions of clusters.

The values of gas mass fraction
obtained for the clusters surrounding the FRII sources studied
here are consistent with observed values for central regions of
clusters, after the correlation between
the radio spectral index and redshift is taken into account, 
and a redshift evolution of the cluster core radius is
considered.  
This suggests that the cluster core radius evolution and 
the effects of the radio spectral index-redshift correlation are
important. 

The gas mass fraction seems to decrease slowly with redshift.  
The increase of cluster core radius with redshift can be one
cause. Another possible explanation is that 
a large fraction of the gas is still in the galaxies
at high redshift, and is later stripped from the galaxies into the
cluster at low redshift. Cooling of the cluster gas also tends to
cause the gas mass fraction in the cluster center to increase with time.

The results on mass and gas fraction should be taken as
preliminary because of the small size of the sample.

\acknowledgments

It is a pleasure to thank Greg Wellman for 
helpful discussions.  
This work was supported in part by 
the US National Science Foundation, the Independent College Fund of New
Jersey, and by a grant from W. M. Wheeler III.

\clearpage

\figcaption[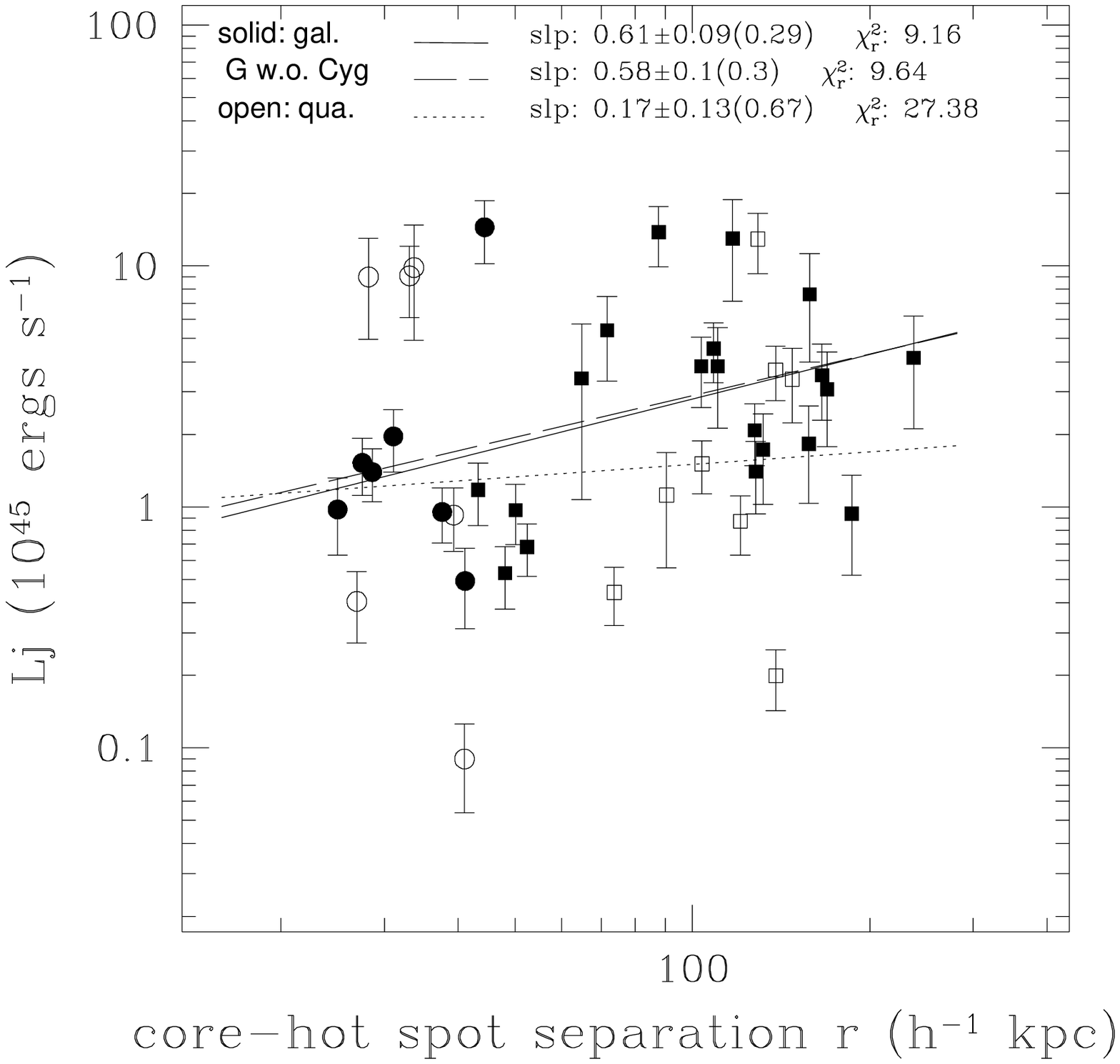]{The luminosity in directed kinetic energy of the jet vs.
  the core-hot spot separation. The best-fitted
  lines for all galaxies, all galaxies except Cygnus A, and all
  radio-loud quasars, are shown with a solid line, a dashed line
  and a dotted line, respectively. Slopes of the best-fitted
  lines are labeled. The number in parenthesis is the uncertainty
  on the slope times $\sqrt{r\chi^2}$, which includes the
  effect of a reduced $\chi^2$, defined as $r\chi^2$,
  greater than one for the fit.}

\figcaption[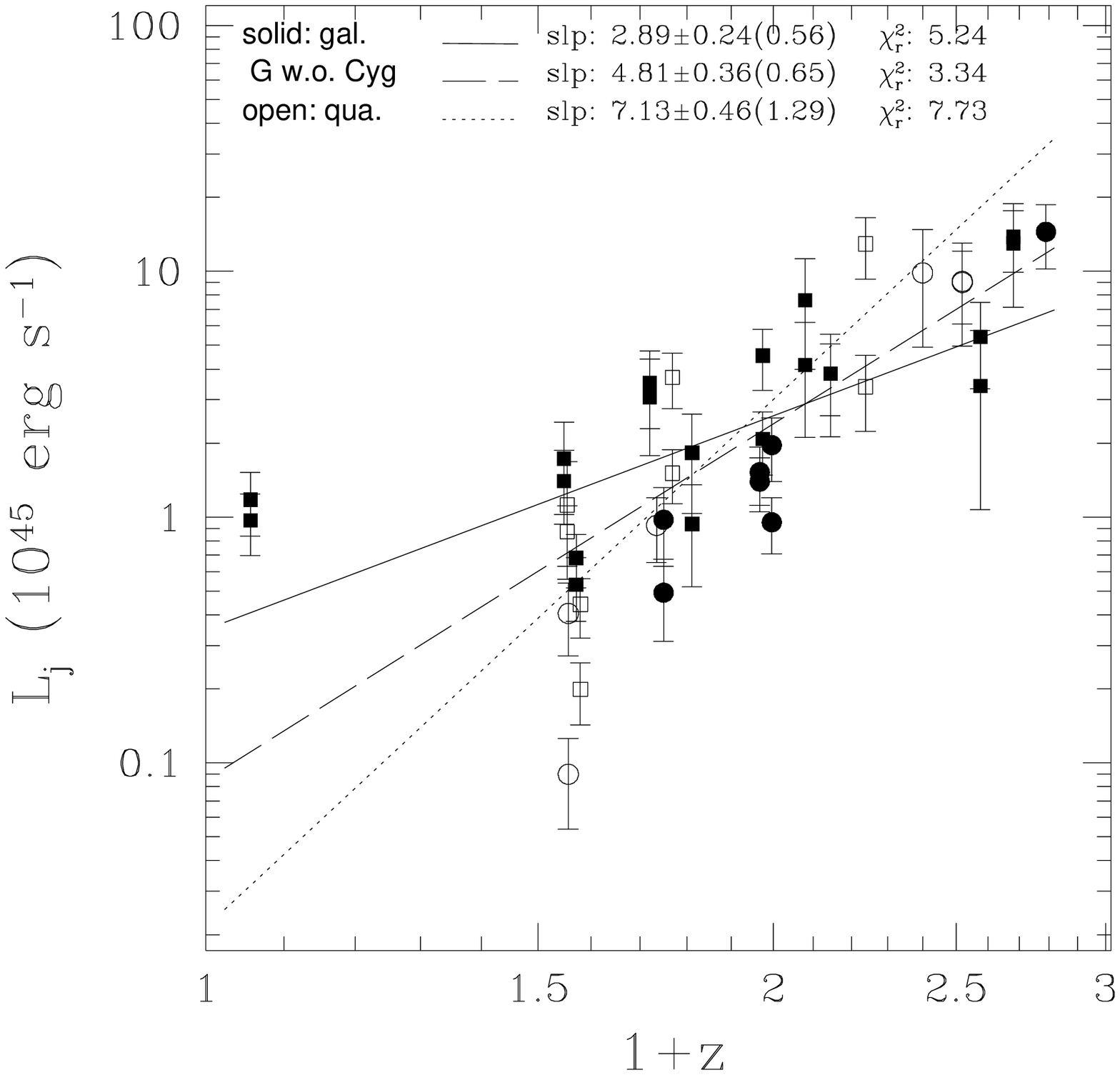]{The luminosity in directed kinetic
  energy of the jet vs. redshift.}

\figcaption[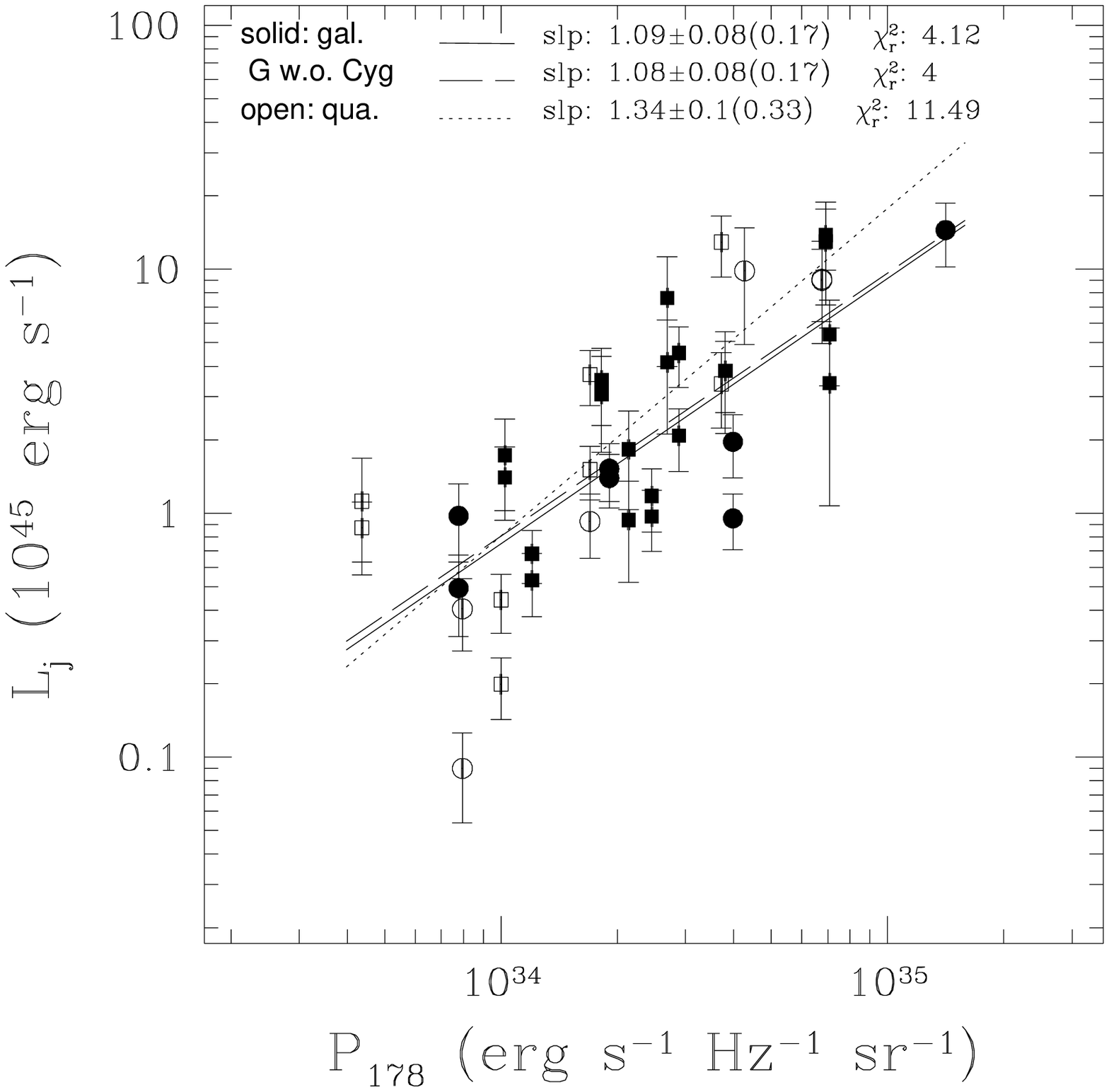]{The luminosity in directed kinetic
  energy of the jet vs. the radio power at 178 MHz.}
 
\figcaption[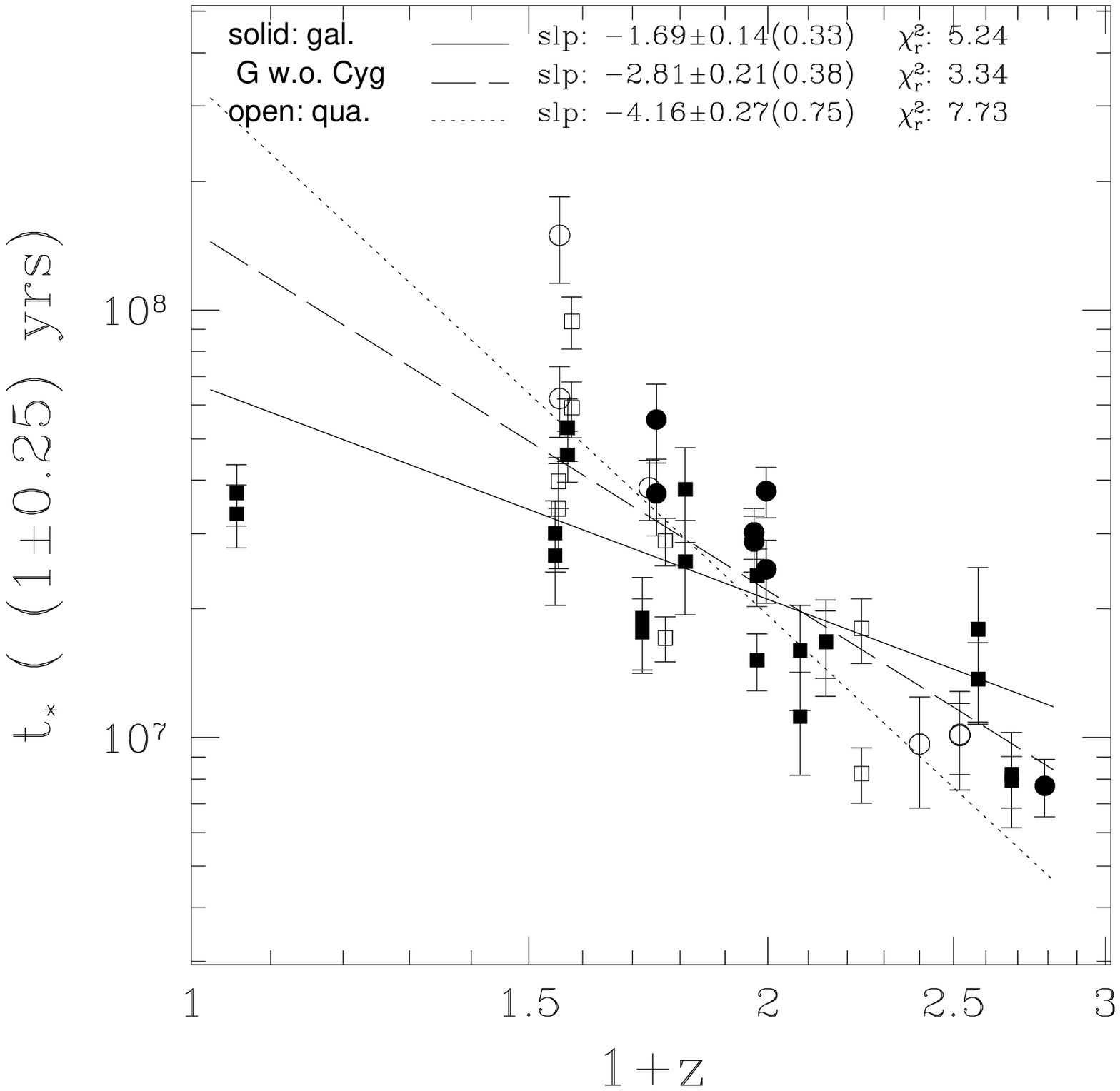]{Lifetime of the collimated outflows vs. redshift.}
 
\figcaption[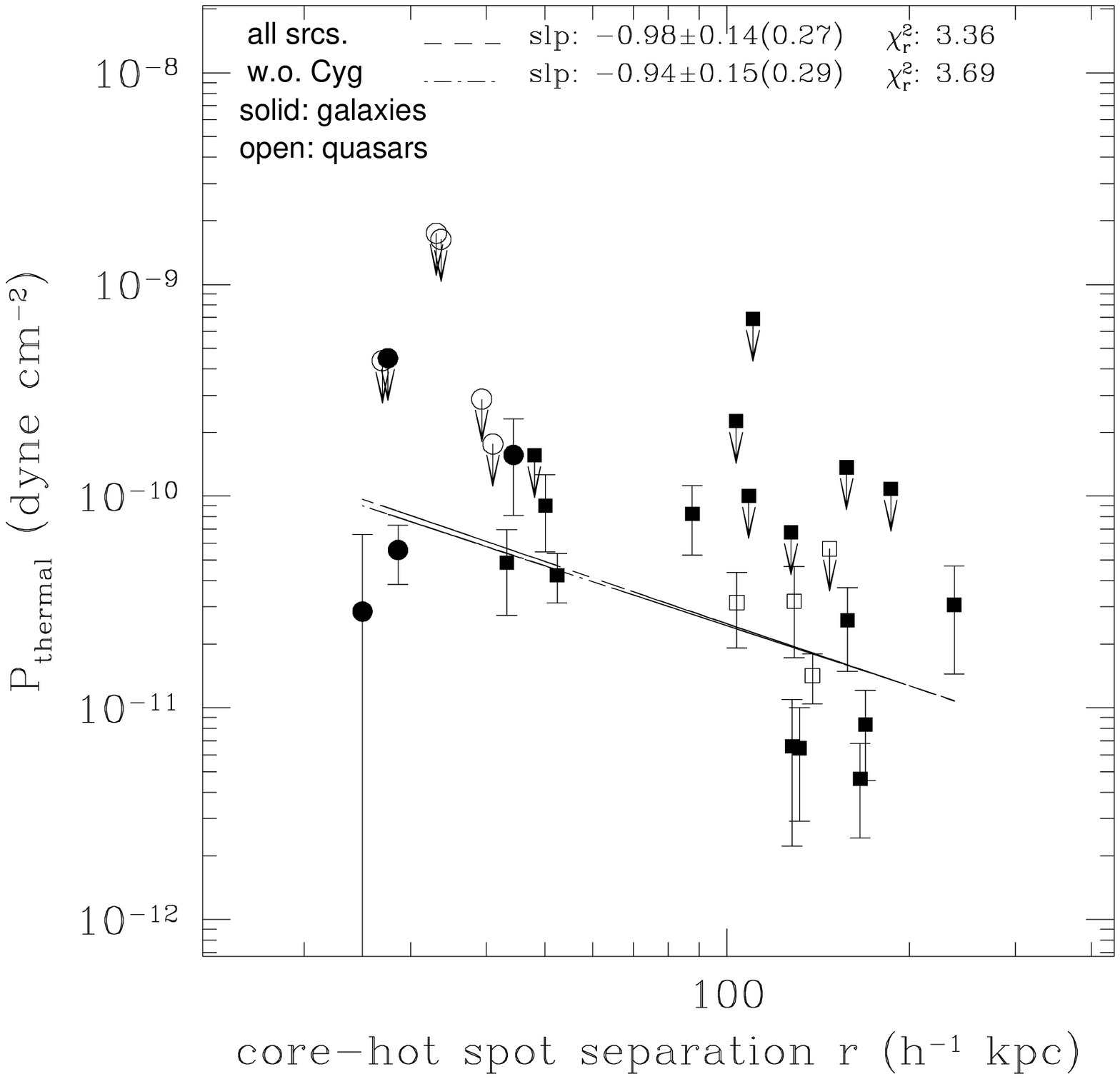]{Thermal pressure of the ambient gas in the vicinity of the
  radio lobe vs. the core-hot spot separation.  The best-fitted
  lines for all sources, including galaxies and quasars,
  and all sources except Cygnus~A are shown with a dashed 
  line and a dash-dot line, respectively. Only sources with
  detections on the Mach number $M$, and hence $P_{th}$, 
  are included in the fit.  }

\figcaption[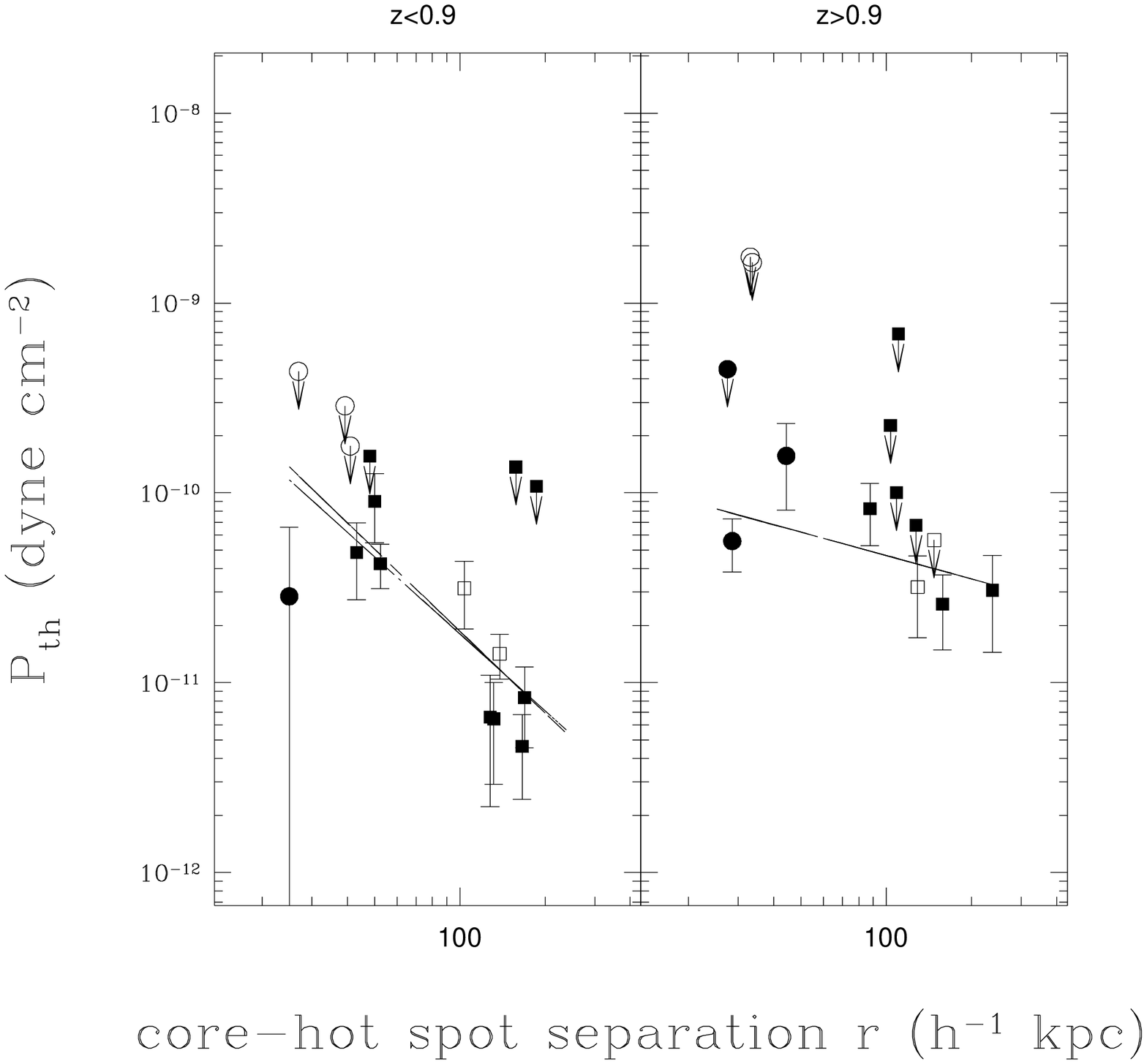]{Thermal pressure of the ambient gas in the vicinity of the
  radio lobe vs. the core-hot spot separation for two redshift
  bins ($z<0.9$, $z>0.9$). In the low-redshift bin, slopes of the best-fitted
  lines are $-1.43\pm0.21(0.27)$ with a reduced $\chi^2$ of 1.58
  for all sources, and $-1.35\pm0.25(0.32)$ with a reduced
  $\chi^2$ of 1.65 for all sources except Cygnus~A. 
  In the high redshift bin, the slope of the best-fitted
  line is $-0.41\pm0.20(0.30)$ with a reduced $\chi^2$ of 2.29.
  Only sources with  detections on $P_{th}$ are included in the fit. }

\figcaption[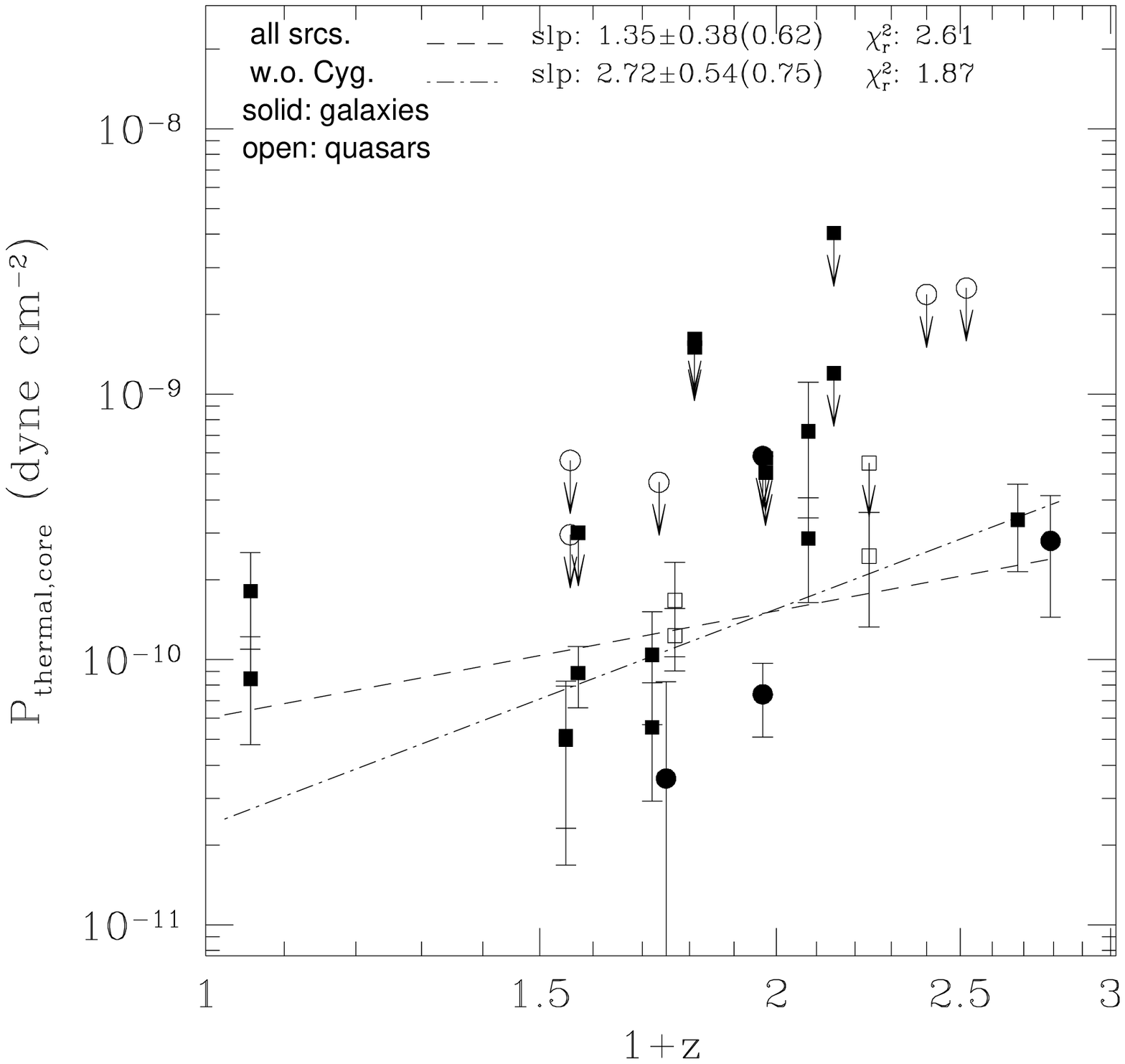]{Thermal pressure of the ambient gas at the cluster center
  vs. redshift for an isothermal gas distribution that follows
  the King density profile. A non-evolving cluster radius of 
  $50 h^{-1}$ kpc, and $\beta_0=2/3$ are used. The same plots for an 
  evolving core radius of $50 (1+z)^{1.6} h^{-1}$ kpc are very
  similar to these and are not shown here. Bounds 
  are not included in the fits. }

\figcaption[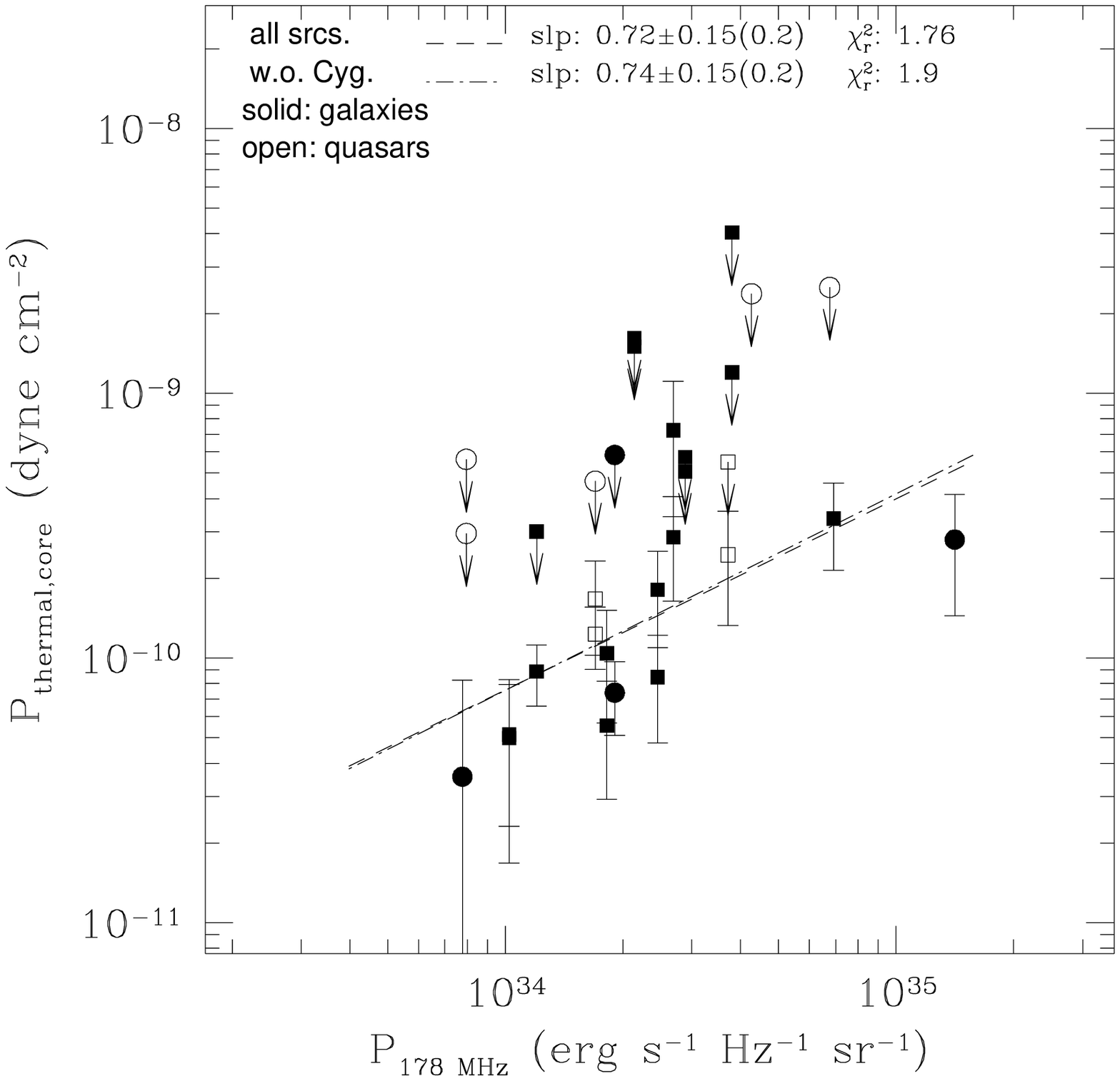]{Thermal pressure of the ambient gas at the cluster center
  vs. the radio power at 178 MHz for an isothermal gas distribution that follows
  the King density profile. A non-evolving cluster radius of 
  $50 h^{-1}$ kpc, and $\beta_0=2/3$ are used.  The same plots for an 
  evolving core radius of $50 (1+z)^{1.6} h^{-1}$ kpc are very
  similar to these and are not shown here. Bounds 
  are not included in the fits.   }

\figcaption[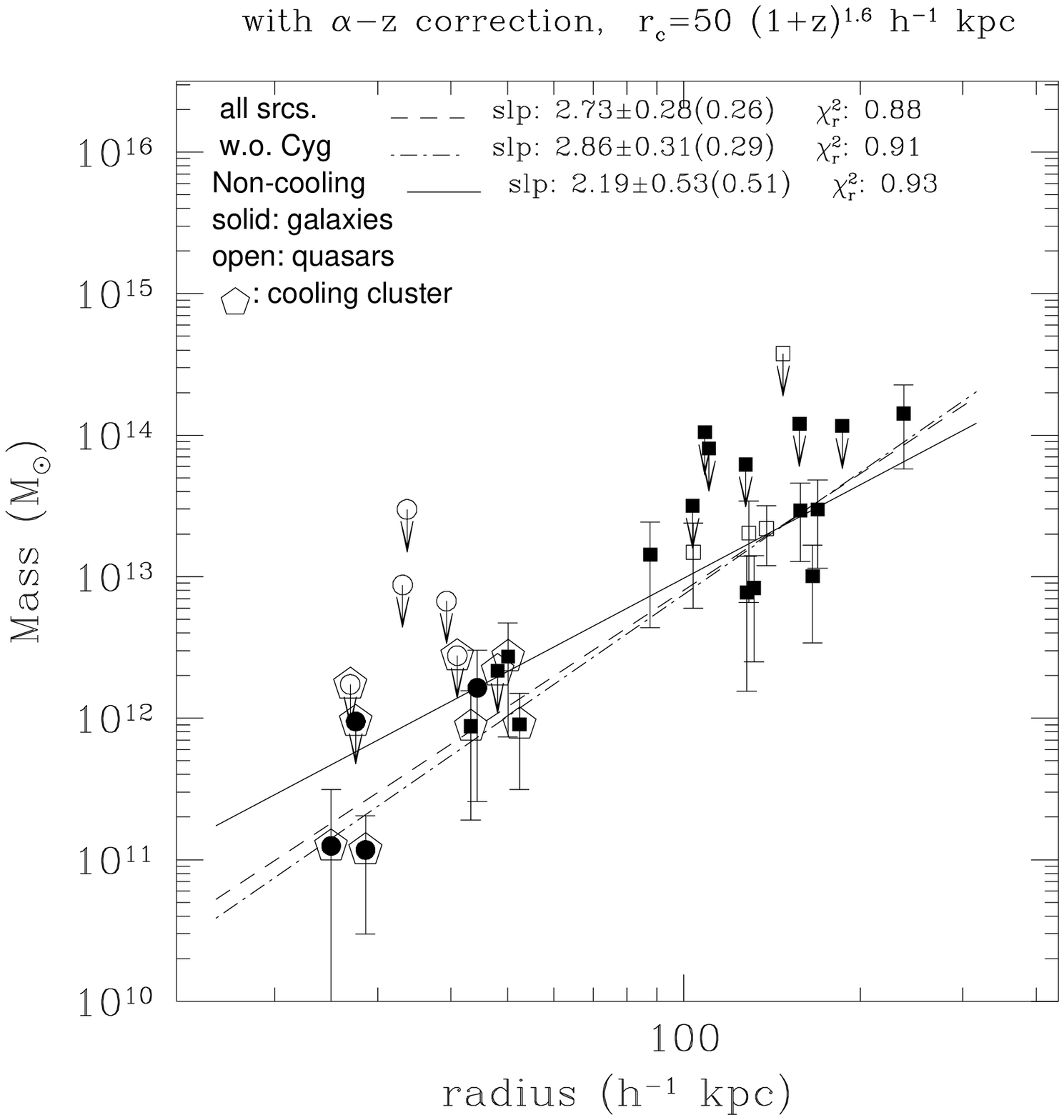]{Cluster mass within $r$ vs. the core-hot spot 
  separation $r$. 
  An isothermal gas distribution that follows the King density profile
  with an evolving cluster core radius of \hbox{$50\, (1+z)^{1.6} h^{-1}$ kpc}
  and $\beta_0=2/3$ is used. The radio spectral index is
  redshift-corrected. Sources that are
  cooling at the point where the cluster temperature $T$ is measured
  are marked with pentagons. The
  best-fitted lines for all sources, including both galaxies and
  quasars, all sources except Cygnus~A, and all sources that are
  not cooling, are shown with a dashed line, a dot-dash line, and
  a solid line, respectively. Only sources with detections on
  $T$, hence mass, are included in the fit.  }

\figcaption[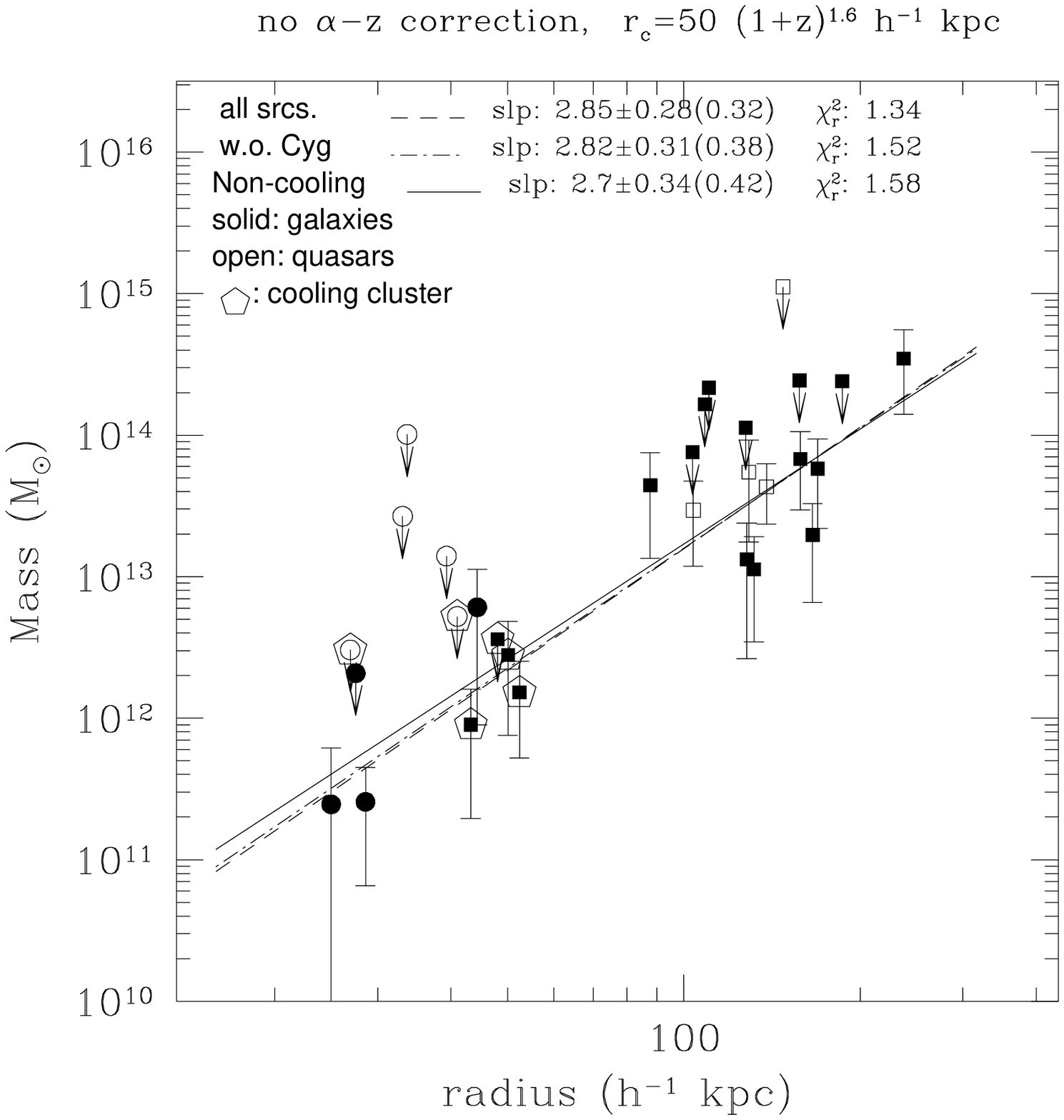]{Cluster mass within $r$ vs. the core-hot spot 
  separation $r$. 
  An isothermal gas distribution that follows the King density profile
  with an evolving cluster core radius of \hbox{$50\, (1+z)^{1.6} h^{-1}$ kpc}
  and $\beta_0=2/3$ is used. 
  The radio spectral index is not redshift-corrected. }

\figcaption[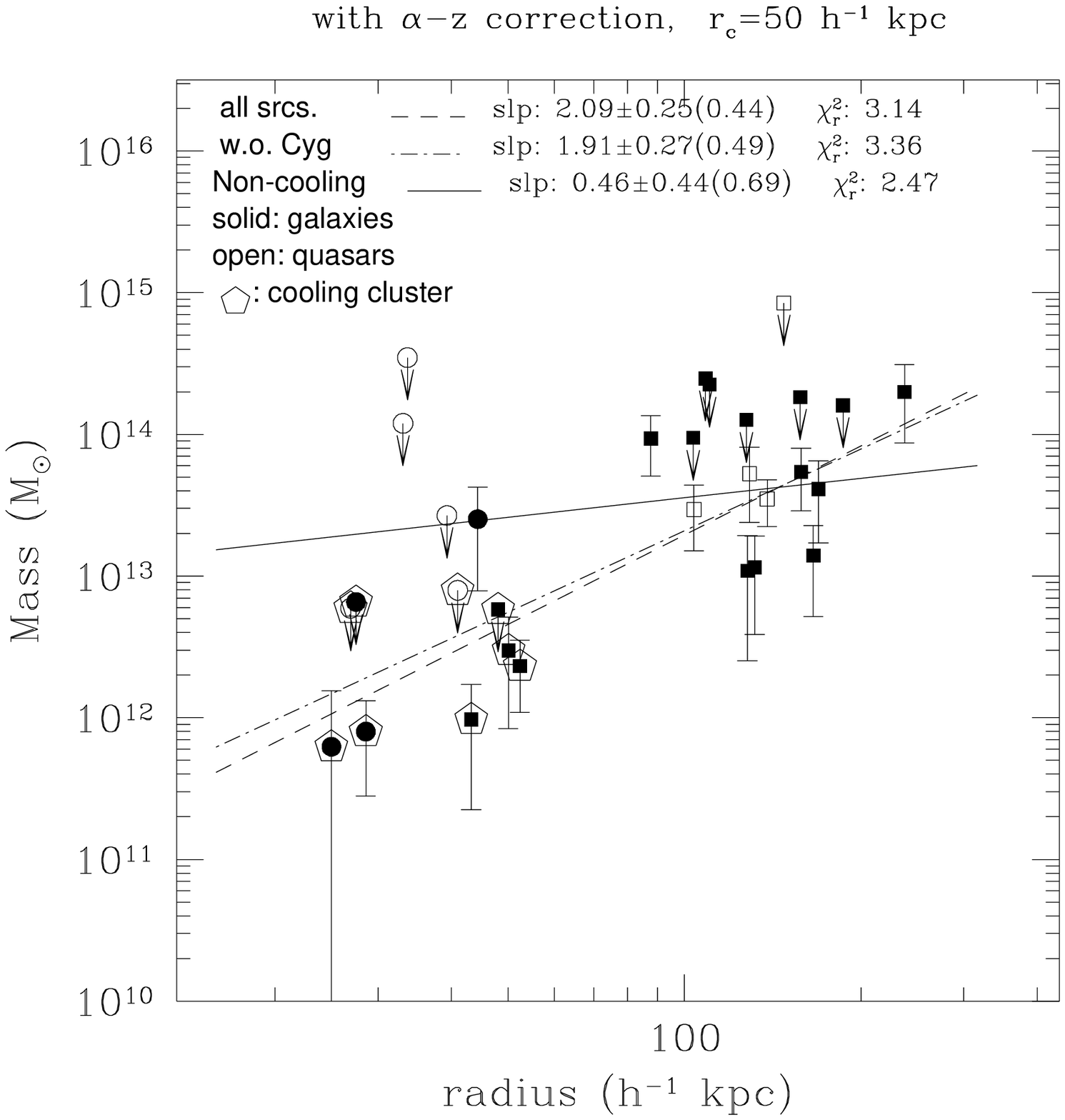]{Cluster mass within $r$ vs. the core-hot spot separation $r$.
  An isothermal gas distribution that follows the King density profile
  with a fixed cluster core radius of \hbox{$50\, h^{-1}$ kpc}
  and $\beta_0=2/3$ is used. 
  The radio spectral index is redshift-corrected.}

\figcaption[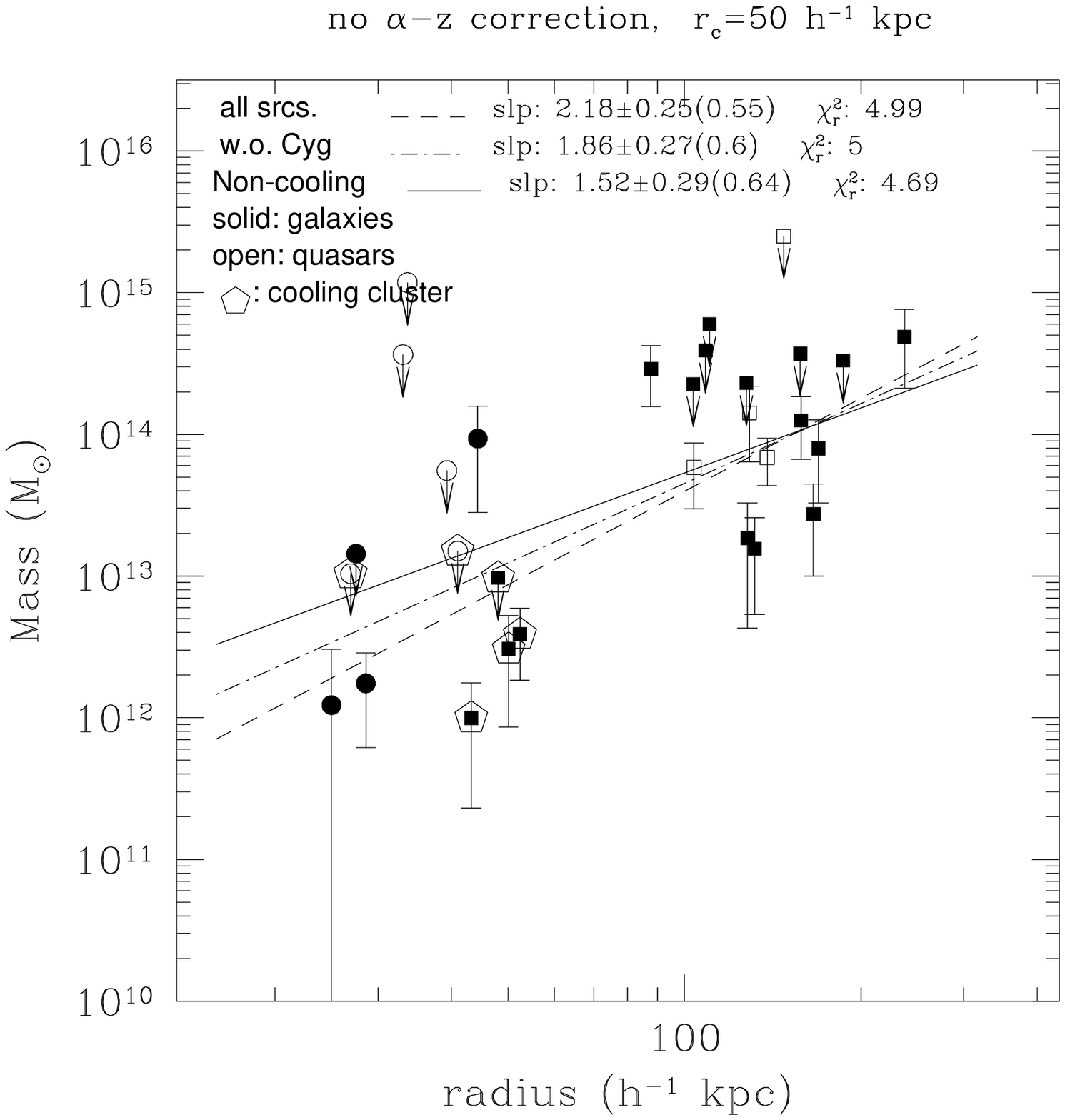]{Cluster mass within $r$ vs. the core-hot 
  spot separation $r$.
  An isothermal gas distribution that follows the King density profile
  with a fixed cluster core radius of \hbox{$50\, h^{-1}$ kpc}
  and $\beta_0=2/3$ is used. 
  The radio spectral index is not redshift-corrected. }

\figcaption[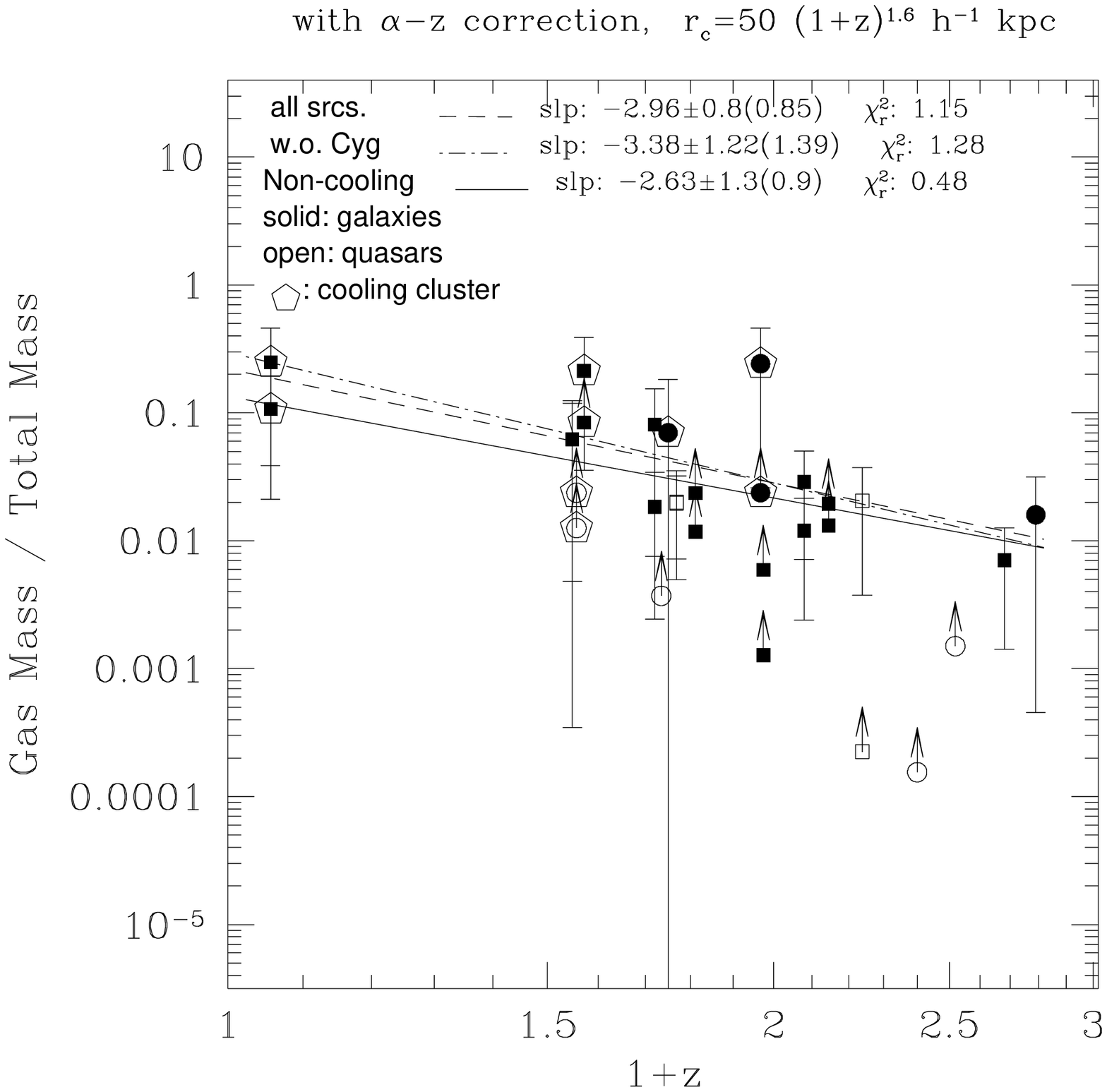]{Gas fraction ($M_{gas}/M_{t}$) within radius $r$ 
  vs. redshift. 
  The radio spectral index is redshift-corrected, and
  an isothermal gas distribution that follows the King density profile
  with an evolving cluster core radius of \hbox{$50\, (1+z)^{1.6} h^{-1}$ kpc}
  and $\beta_0=2/3$ is used. }


\clearpage

\begin{table}
\dummytable      
\label{tab:data}
\end{table}

\begin{deluxetable}{lrllrrrr}
 \small
\tablecaption{\label{tab:ma3p}Results Obtained by Fitting
$M_t(r) \propto r^{n_1} (1+z)^{n_2} P_{178}^{n_3}$}
\tablehead{
\colhead{type\tablenotemark{a}}	&
\colhead{No.\tablenotemark{b}}	&
\colhead{$\alpha$-$z$?\tablenotemark{c}}	&
\colhead{$r_c$-$z$?\tablenotemark{d}}	&
\colhead{$n_1$\tablenotemark{e}}	&
\colhead{$n_2$\tablenotemark{f}}	&
\colhead{$n_3$\tablenotemark{g}}  &
\colhead{$r\chi^2$\tablenotemark{h}} }
\startdata
 %
 N.C.Cl.  &  11  &  Yes  &  Yes  &  2.38 $\pm$ 0.70(0.52)  &   6.86 $\pm$ 5.48(4.1)  & -1.31 $\pm$ 1.57(1.17)  &  0.56  \nl
   G+Q-C  &  14  &  Yes  &  Yes  &  2.95 $\pm$ 0.31(0.23)  &   2.72 $\pm$ 3.80(2.84) & -0.08 $\pm$ 1.01(0.76)  &  0.56  \nl
     G+Q  &  16  &  Yes  &  Yes  &  2.88 $\pm$ 0.30(0.23)  &  -0.73 $\pm$ 0.96(0.73)  & 0.80 $\pm$ 0.37(0.28)  &  0.58  \nl
\tablevspace{5pt} \hline \tablevspace{5pt}
 N.C.Cl.  &  13  &   No  &  Yes  &  3.05 $\pm$ 0.36(0.28)  &  4.46 $\pm$ 4.11(3.18)  & -0.13 $\pm$ 1.06(0.82)  &  0.60  \nl
   G+Q-C  &  14  &   No  &  Yes  &  2.95 $\pm$ 0.31(0.23)  &  3.66 $\pm$ 3.82(2.88)  &  0.02 $\pm$ 1.02(0.77)  &  0.57  \nl
     G+Q  &  16  &   No  &  Yes  &  2.89 $\pm$ 0.30(0.23)  &  0.49 $\pm$ 0.96(0.73)  &  0.83 $\pm$ 0.37(0.28)  &  0.58  \nl
\tablevspace{5pt} \hline \tablevspace{5pt}
 N.C.Cl.  &  11  &  Yes  &   No  &  1.36 $\pm$ 0.63(0.52)  &  7.40 $\pm$ 4.55(3.72)  & -0.93 $\pm$ 1.33(1.09)  &  0.67  \nl
   G+Q-C  &  14  &  Yes  &   No  &  2.12 $\pm$ 0.27(0.23)  &  3.18 $\pm$ 3.33(2.85)  &  0.44 $\pm$ 0.90(0.77)  &  0.73  \nl
     G+Q  &  16  &  Yes  &   No  &  2.11 $\pm$ 0.26(0.21)  &  1.63 $\pm$ 0.91(0.75)  &  0.84 $\pm$ 0.33(0.27)  &  0.68  \nl
\tablevspace{5pt} \hline \tablevspace{5pt}
 N.C.Cl.  &  13  &   No  &   No  &  2.17 $\pm$ 0.31(0.28)  &  4.45 $\pm$ 3.40(3.10)  &  0.47 $\pm$ 0.88(0.80)  &  0.83  \nl
   G+Q-C  &  14  &   No  &   No  &  2.12 $\pm$ 0.27(0.23)  &  4.09 $\pm$ 3.20(2.77)  &  0.53 $\pm$ 0.86(0.74)  &  0.75  \nl
     G+Q  &  16  &   No  &   No  &  2.11 $\pm$ 0.27(0.22)  &  2.84 $\pm$ 0.91(0.76)  &  0.86 $\pm$ 0.33(0.27)  &  0.69  \nl
\enddata
\tablenotetext{a}{Type of sources included in the fit:  ``G+Q'' refers
to all sources with temperature estimates, including radio
galaxies and radio-loud quasars, ``G+Q-C'' refers to sources other than
Cygnus~A, and ``N.C.Cl.'' refers to clusters that are not cooling
at the position where the temperature is measured.}
\tablenotetext{b}{Number of data points used for the fit.}
\tablenotetext{c}{Whether the redshift-correction on the radio spectral
index is applied.}
\tablenotetext{d}{Whether a redshift evolution of the cluster
core radius $r_c$ is considered. The core radius is taken to be
\hbox{$50\,(1+z)^{1.6} h^{-1}\,\mbox{kpc}$} if an evolution is
considered, and \hbox{$50\,h^{-1}\,\mbox{kpc}$} if no evolution is
considered (see \S\ref{ssec:pth-r}).}
\tablenotetext{e}{$n_1$ and its error. The number in parenthesis
is the error on $n_1$ times $\sqrt{r\chi^2}$ which includes the
effect of the reduced $\chi^2$, defined as $r\chi^2$, of the fit.}
\tablenotetext{f,g}{same as note c for $n_2$ and $n_3$, respectively.}
\tablenotetext{h}{Reduced $\chi^2$ of the fit.}
\end{deluxetable}

\begin{deluxetable}{lrllrrr}
 \small
\tablecaption{\label{tab:ma2p}Results Obtained by Fitting
$M_t(r) \propto r^{n_1} (1+z)^{n_2}$}
\tablehead{
\colhead{type\tablenotemark{a}}	&
\colhead{No.\tablenotemark{b}}	&
\colhead{$\alpha$-$z$?\tablenotemark{c}}	&
\colhead{$r_c$-$z$?\tablenotemark{d}}	&
\colhead{$n_1$\tablenotemark{e}}	&
\colhead{$n_2$\tablenotemark{f}}	&
\colhead{$r\chi^2$} }
\startdata
 %
 N.C.Cl.  &  11  &  Yes  &  Yes  &  2.70 $\pm$ 0.59(0.45)  &  2.39 $\pm$ 1.24(0.94)  &  0.58  \nl
   G+Q-C  &  14  &  Yes  &  Yes  &  2.95 $\pm$ 0.31(0.22)  &  2.43 $\pm$ 1.05(0.75)  &  0.51  \nl
     G+Q  &  16  &  Yes  &  Yes  &  2.67 $\pm$ 0.29(0.28)  &  0.63 $\pm$ 0.73(0.69)  &  0.90  \nl
\tablevspace{5pt} \hline \tablevspace{5pt}
 N.C.Cl.  &  13  &   No  &  Yes  &  3.04 $\pm$ 0.35(0.26)  &  3.99 $\pm$ 1.16(0.86)  &  0.55  \nl
   G+Q-C  &  14  &   No  &  Yes  &  2.95 $\pm$ 0.31(0.22)  &  3.74 $\pm$ 1.06(0.76)  &  0.52  \nl
     G+Q  &  16  &   No  &  Yes  &  2.67 $\pm$ 0.29(0.28)  &  1.90 $\pm$ 0.73(0.70)  &  0.93  \nl
\tablevspace{5pt} \hline \tablevspace{5pt}
 N.C.Cl.  &  11  &  Yes  &   No  &  1.60 $\pm$ 0.52(0.42)  &  4.30 $\pm$ 1.04(0.84)  &  0.65  \nl
   G+Q-C  &  14  &  Yes  &   No  &  2.13 $\pm$ 0.27(0.22)  &  4.75 $\pm$ 0.83(0.68)  &  0.68  \nl
     G+Q  &  16  &  Yes  &   No  &  1.88 $\pm$ 0.25(0.27)  &  3.34 $\pm$ 0.62(0.66)  &  1.13  \nl
\tablevspace{5pt} \hline \tablevspace{5pt}
 N.C.Cl.  &  13  &   No  &   No  &  2.20 $\pm$ 0.31(0.27)  &  6.19 $\pm$ 0.94(0.82)  &  0.77  \nl
   G+Q-C  &  14  &   No  &   No  &  2.13 $\pm$ 0.27(0.23)  &  6.00 $\pm$ 0.83(0.70)  &  0.72  \nl
     G+Q  &  16  &   No  &   No  &  1.88 $\pm$ 0.25(0.27)  &  4.57 $\pm$ 0.62(0.67)  &  1.16  \nl
\enddata
\tablenotetext{a-f}{same as those in Table~\ref{tab:ma3p}.}
\end{deluxetable}

\begin{figure}
 \label{fig:ljr}
\end{figure}

\begin{figure}
 \label{fig:ljz}
\end{figure}

\begin{figure}
 \label{fig:ljp}
\end{figure}

\begin{figure}
 \label{fig:tstarz}
\end{figure}

\begin{figure}
 \label{fig:pthr}
\end{figure}

\begin{figure}
 \label{fig:pthrhilo}
\end{figure}

\begin{figure}
 \label{fig:pthcz}
\end{figure}

\begin{figure}
 \label{fig:pthcp}
\end{figure}

\begin{figure}
 \label{fig:mtr42e}
\end{figure}

\begin{figure}
 \label{fig:mtr02e}
\end{figure}

\begin{figure}
 \label{fig:mtr42}
\end{figure}

\begin{figure}
 \label{fig:mtr02}
\end{figure}

\begin{figure}
 \label{fig:fg42e}
\end{figure}

\end{document}